\documentclass{aa}  

\usepackage{graphicx}
\usepackage{txfonts}
\usepackage{scrextend}
\usepackage{hyperref}
\begin{document} 

\newcommand*{\mercury}{\textsc{mercury6}\xspace}
\newcommand*{\horizons}{\textsc{horizons}\xspace}

   \title{Towards an explanation of orbits in the extreme trans-Neptunian region:
   The effect of Milgromian dynamics}

   \author{R. Pau\v{c}o
          }

   \institute{Faculty of Mathematics, Physics and Informatics, Comenius University in Bratislava,
    Mlynsk\'{a} dolina, 842 48 Bratislava\\
              \email{pauco@fmph.uniba.sk}
             }

   \date{received december 27, 2016; accepted march 16, 2017}

 
\abstract
{Milgromian dynamics (MD or MOND) uniquely predicts motion in a galaxy from the distribution of its stars and gas in a remarkable agreement with observations so far. In the solar system, MD predicts the existence of some possibly non-negligible dynamical effects, which can be used to constrain the freedom in MD theories. Known extreme trans-Neptunian objects (ETNOs) have their argument
of perihelion, longitude of ascending node, and inclination distributed in
highly non-uniform fashion; ETNOs are bodies with perihelion distances greater than the orbit of Neptune and with semimajor axes greater than 150 au and less than $\sim1500$ au. It is as if these bodies have been systematically perturbed by some external force.}
{We investigated a hypothesis that the puzzling orbital characteristics of ETNOs are a consequence of MD.}
{We set up a dynamical model of the solar system incorporating the external field effect (EFE), which is anticipated to be the dominant effect of MD in the ETNOs region. We used constraints available on the strength of EFE coming from radio tracking of the Cassini spacecraft. We performed several numerical experiments, concentrating on the long-term orbital evolution of primordial (randomised) ETNOs in MD.} 
{The EFE could produce distinct non-uniform distributions of the orbital elements of ETNOs that are related to the orientation of an orbit in space. If we demand that EFE is solely responsible for the detachment of Sedna and 2012 VP$_{113}$, then these distributions are at odds with the currently observed statistics on ETNOs unless the EFE quadrupole strength parameter $Q_{2}$ has values that are unlikely (with probability < 1$\%$) in light of the Cassini data.}
{}

   \keywords{minor planets, asteroids: general --- gravitation --- Kuiper belt: general}

   \maketitle
%

\section{Introduction}

History has taught us that we should pay attention to orbital anomalies in the solar system. 
Two of the best historical examples are the discovery of Neptune and nonexistence of the planet Vulcan.
Neptune was found right where it was predicted to erase an anomaly in the orbit of Uranus \citep{LeV46}, while the non-existence of Vulcan finally led to the development of new revolutionary ideas in physics -- the geometrical interpretation of gravity -- explaining precisely the minute anomalous precession of Mercury \citep{LeV59,E1916}

Orbital characteristics of observed extreme trans-Neptunian objects (ETNOs), objects having perihelion distances, $q$, greater than the orbit of Neptune, and semimajor axes, $a$, greater than 150 au and less than roughly 1500 au (definition established and motivated in \citealp{TS14}), suggest that these objects have been systematically perturbed by some external force.
On September 1, 2016, the Minor Planet Center (MPC) orbit database registered 19 ETNOs\footnote{All ETNOs have $q>33$ au, 14 of 19 ETNOs have $q>35$ au, and 8 of 19 have $q>40$ au.}. 
The outlying positions of 90377 Sedna and 2012 VP$_{113}$ (hereafter, collectively referred to as Sednoids) in the $q$ versus eccentricity, $e$, diagram (e.g. Fig. 2 in \citealp{ST16}) cannot be addressed by the interaction with the known solar system planets and the current Galactic environment (e.g. \citealp{D+15}). Nevertheless, orbits of ETNOs are also characterised by argument of perihelion, $\omega$, longitude of the ascending node, $\Omega$, and perihelion longitude, $L$,\footnote{Longitude of an object at perihelion in the heliocentric ecliptic coordinates. It is not equal to the longitude of perihelion, $\varpi\equiv \omega+\Omega$, except for the case of nil inclination. Similarly, when we refer to perihelion latitude, $B$, we mean latitude of an object at perihelion in the heliocentric ecliptic coordinates. Orbital elements $\omega$, $\Omega$, and $i$ give $L$ and $B$ unambiguously: $\sin (L-\Omega) = \sin\omega\cos i / \cos B$, $\cos(L-\Omega)=\cos\omega/\cos B$.} which are distributed in a highly non-uniform and clustered fashion (\citealp{TS14}, \citealp{dlFdlF14}, \citealp{BF16}, \citealp{BB16}). This is an unexpected observation, as these orbital elements are expected to circulate rapidly under the action of the giant planets.

Tightly clustered are also the eccentricity, $e$ ($e=0.82\pm 0.06$), and inclination, $i$ ($i=20\pm 7$ deg) of the ETNOs, where in brackets we present mean $\pm 1\sigma$ values. 
The observed distribution of eccentricities is the result of a simple bias. We preferentially find objects close to the Earth, hence objects with smaller perihelia and hence larger eccentricities are preferred. Such bias means that most objects should be found with eccentricities in the range 0.8 -- 0.9 \citep{dlFdlF14}.
However, the clustering in $i$ should be of dynamical origin as the intrinsic bias in declination owing to our observations made from the Earth, prefers lower inclinations\footnote{We add to this that many surveys are performed preferentially close to the ecliptic.} \citep{dlFdlF14}.

\cite{BB16} recognised that orbits of long-periodic ETNOs with $a>250$ au (six such objects were known at that time\footnote{Asteroid 2000 CR$_{105}$ with $a=228$ au is very close to the six in terms of $\omega$, $\Omega$, and $L$, however, asteroid 2001 FP$_{185}$, with $a=226$ au, is about 90 deg ahead in $L$. Hence, the boundary was identified to be at $a=250$ au.}: Sedna, 2004 VN$_{112}$, 2007 TG$_{422}$, 2010 GB$_{174}$, 2012 VP$_{113}$, and 2013 RF$_{98}$) 
are very tightly clustered in terms of $\omega$, $\Omega$, and $L$, and hence, aligned in physical space.
This observational fact, plus the large perihelia of Sednoids, has led \citet{BB16} (see also \citealp{BB16b}) to hypothesise the existence of the ninth planet of the solar system, dubbed Planet 9 (hereafter P9), with mass of $\sim 10$ $M_{\oplus}$, perihelion longitude 180 deg ahead of the six, orbiting in an eccentric, $a\sim 700$ au, $e\sim 0.6$, and moderately inclined orbit. Discussion in the astronomical community settled in a consensus that the orbital characteristics of ETNOs are not result of an observational bias and/or smallness of the sample, and the quest of finding P9 was prompted \citep{BB16b,Cow+16,dlFdlF16,dlF+16,For+16,Fie+16,HP16,HP16b}. 
Recently, \cite{ST16} found four new ETNOs, two of which have $a>250$ au (2014 SR$_{349}$ and 2013 FT$_{28}$). These two ETNOs have $\omega$ and $i$ clustered in line with the original six, but one of them, 2013 FT$_{28}$, has $\Omega$ and $L$ with 90 -- 180 deg offset from the original clustering positions.

\cite{S+16} demonstrated a drawback of the P9 scenario; they showed that P9 generically drives ETNOs into giant planets crossing orbits, which subsequently decouples them from any shepherding mechanism of P9 and randomises their $\omega$, $\Omega$, and $L$. Hence, these randomised ETNOs should contaminate the observed sample. However, no such population is observed.
In \cite{Law+17}, the authors evolved primordial planetesimal disk for 4 Gyr in the solar system with the present-day planetary architecture plus an additional 10 $M_{\oplus}$ planet. Both circular \citep{TS14} and eccentric \citep{BB16} orbits of the ninth planet were considered. They found no evidence for orbital clustering in the ETNOs region at the end of the simulation neither for a circular nor an eccentric ninth planet. These findings bring some controversy into the P9 scenario.

Knowing the distribution of stars and gas (baryons), the matter we see, the motion in galaxies is determinable. This is implied by the existence of the Galactic laws, such as the baryonic Tully-Fisher relation \citep{M+00}, the relation between stellar and dynamical central surface densities \citep{L+16}, and the mass discrepancy-acceleration correlation \citep{San90,McG04,Des16}. All these can be encapsulated into one law called the radial acceleration relation \citep{McG+16,L+17} -- a very tight correlation between the radial acceleration traced by rotation curves and the acceleration predicted by the observed distribution of baryons. There is no reference to dark matter in these laws, even in systems that are supposedly dark matter dominated, i.e. systems showing large mass-discrepancy in Newtonian dynamics. 
Additionally, ``for any feature in the luminosity profile there is a corresponding feature in the rotation curve and vice versa'' \citep{Ren04}. 
In the context of dark matter halos, the existence of these laws means that there is a very tight correlation between the baryonic and dark matter distributions. Such a very tight correlation, independent of properties and histories of individual galaxies, is unexpected in the current cosmological paradigm: the $\Lambda$CDM model, where $\Lambda$ is the positive cosmological constant representing dark energy, whose nature is unknown, and CDM stands for cold (non-baryonic) dark matter. All these laws were predicted a priori and follow naturally from the basic principles of Milgromian dynamics \citep{Mil14,Mil16b}.

Milgromian dynamics (MD or MOND = modified Newtonian dynamics; \citealp{Mil83b}), also known as scale-invariant dynamics, uniquely predicts how the distribution of stars and gas in a galaxy relates to its dynamics. 
For reviews on MD, see for example the  \cite{FM12}, which is  thorough and technical, or \cite{Mil16}, which contains a discussion on the potential connection of MD to fundamental physics. Milgromian dynamics departs from Newtonian dynamics (and general relativity) at low accelerations with respect to some predefined frame (this could be e.g. the cosmic microwave background rest frame), and the definition of low acceleration is determined by a new constant of nature, $a_{0}$, having units of acceleration. In the low acceleration regime, unlike Newtonian dynamics and general relativity, MD becomes space-timescale invariant \citep{Mil09b}.
While MD was invented as an alternative to the dark matter hypothesis, which lacks direct observational evidence, over the years MD acquired strong empirical support; see \cite{FM12}, which  summarises the ubiquitous presence of $a_{0}$ in the data. 

In some versions of MD (typically in modified gravity versions of MD), the dynamics of a small subsystem, freely falling in the field of a large external system, is affected by the external free-fall acceleration through the so-called external field effect (EFE; \citealp{Mil83b,FM12}). This is the case of, for instance, a dwarf galaxy in the field of a host galaxy, or the solar system falling freely in the field of the Galaxy. The EFE arises from generic non-linearity of equations of MD and from the fact that it is based on acceleration. \citet{McGM13a} used MD to predict velocity dispersions of dwarf satellite galaxies of the Andromeda galaxy. Later, in \citet{McGM13b}, they compared their a priori predictions with newly available data. They reported that, \textit{``The data are in good agreement with our specific predictions for each dwarf made a priori... MOND distinguishes between regimes where the internal field of the dwarf, or the external field of the host, dominates. The data appear to recognise this distinction, which is a unique feature of MOND not explicable in $\Lambda$CDM.''}
The EFE can also explain the declination of rotation curves in the outer parts of galaxies \citep{Hag+16}. In the outer parts, when the internal accelerations are smaller than the external fields (which are $\lesssim a_{0}$), the quasi-Newtonian behaviour is restored \citep{FM12}.

Recently, \cite{McG16} predicted velocity dispersion, $\sigma_{vel}$, of Crater II in MD. Although internal accelerations are very low $(\ll a_{0})$, EFE reduces $\sigma_{vel}$ down to $\sigma_{vel}\approx2$ km s$^{-1}$. For isolated Crater II (EFE is not present e.g. in modified inertia theories of MD) McGaugh obtained $\sigma_{vel}\approx4$ km s$^{-1}$. Significantly higher $\sigma_{vel}$ would mean falsification of MD. In $\Lambda$CDM, higher $\sigma_{vel}$ is expected \citep{MBK+12,McG16}. The resolution was found recently to be $\sigma_{vel}=2.7\pm0.3$ km s$^{-1}$ \citep{C+16}. This nicely demonstrates predictive power of MD in galaxies.

It was recognised that EFE may act even in systems with high internal accelerations \citep{Mil10}.
\citet{PK16} reconsidered the hypothesis of the Oort cloud of comets in the framework of MD. They also discussed that EFE could produce\footnote{This depends strongly on the shape of MD interpolating function and the value of $a_{0}$. These are free parameters in the theory for the time being (they should follow from a more profound theory), but they are constrained by the effect MD produces in the inner solar system \citep{H+16}.} enough torque to deliver primordial solar system bodies to Sedna-like orbits even in the present-day Galactic environment. Here we reassessed this problem and checked whether the perihelia of both Sednoids can be acquired by the action of EFE. We used recent constraints available on the strength of EFE in the ETNOs region coming from the radio tracking of the Cassini spacecraft \citep{BN11,H+14,H+16}.
Assuming that the statistics on the orbital elements of ETNOs are representative of the real population, we also investigated whether the spatial orientation of the orbits of ETNOS can be explained by this effect of MD as well.

We further discuss the observational data on ETNOs in Sect. \ref{sec:data}.
Milgromian dynamics and the dynamical effect it has on the solar system are introduced in Sect. \ref{sec:MOND}.
We present a dynamical model of the solar system obeying laws of MD that are applicable in the ETNOs region in Sect. \ref{sec:model}. The comparison between the P9 scenario in Newtonian dynamics and the MD scenario without an additional planet is made in Sections \ref{sec:model0} and \ref{sec:results} in the context of the Sednoid perihelia and the spatial orientation of the orbits of ETNOs. We conclude with a summary and discussion of our results in Sect. \ref{sec:discuss}.

\section{More on data}\label{sec:data}

\begin{figure}
\begin{center}
\resizebox{\hsize}{!}{\includegraphics{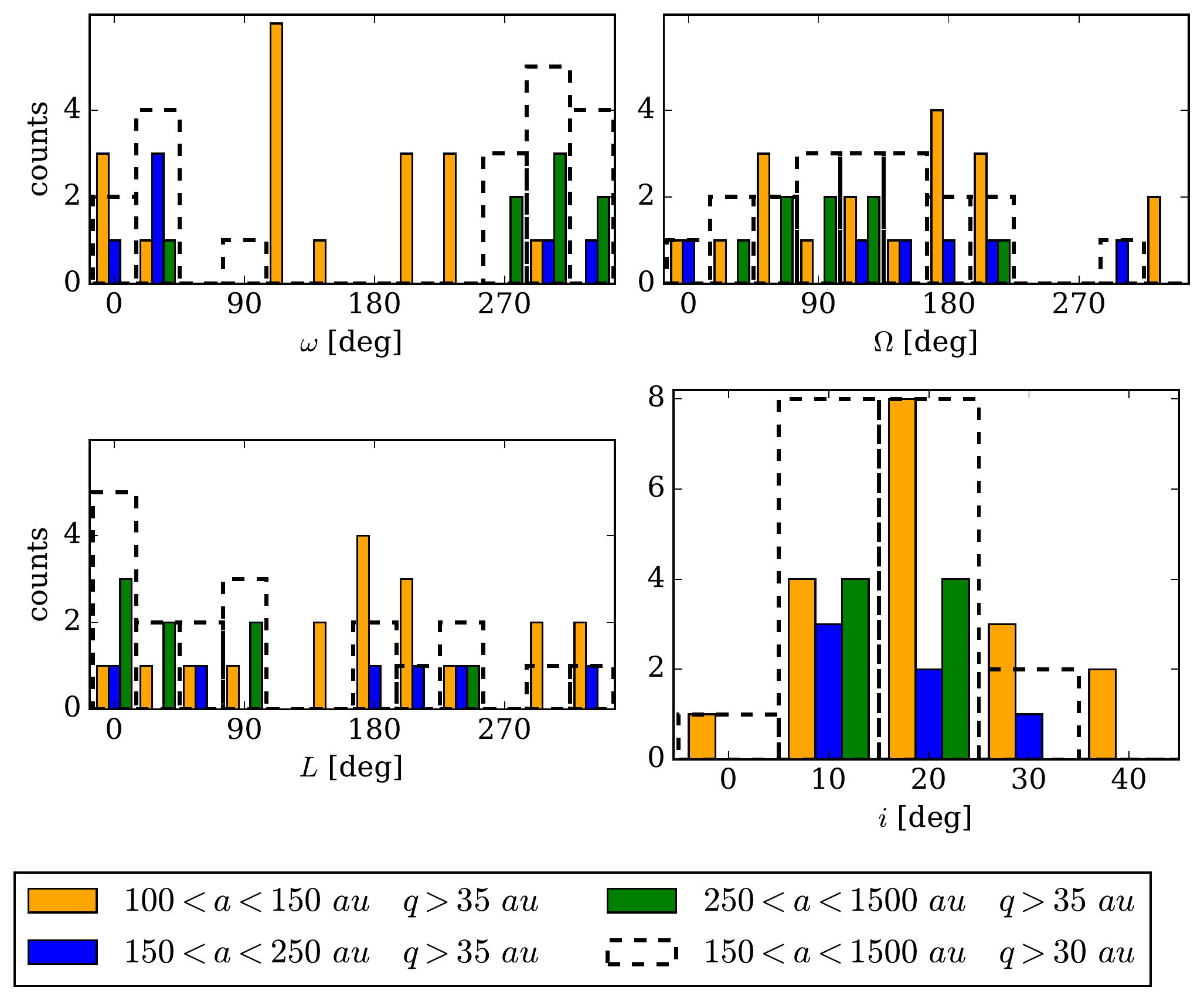}}
\caption{Histogram distributions of argument of perihelion (top left panel), longitude of ascending node (top right panel), perihelion longitude (bottom left panel), and inclination (bottom right panel) inferred from the well-determined orbits of TNOs fulfilling the criteria $100<a<1500$ au, $q>35$ au (colour histograms).
The orbits were discerned according to their $a$ and divided into 3 groups: $100<a<150$ au (yellow histogram), $150<a<250$ au (blue histogram), and $250<a<1500$ au (green histogram). The distributions inferred from all well-determined orbits of ETNOs ($150<a<1500$ au, $q>30$ au) are indicated as dashed bars.
The blue and green bars in a given bin can, but do not have to, sum up to the height of a dashed bar in that bin, as the dashed bar also comprises objects with $q$ in the range from 30 to 35 au. The histogram bars are aligned to the left, for example, in the bottom right panel, 0 indicates inclination range $(0,10)$ deg, 10 indicates inclination range $(10,20)$ deg, and so on.}
\label{img:data1}
\end{center}
\end{figure}

\begin{figure}
\begin{center}
\resizebox{\hsize}{!}{\includegraphics{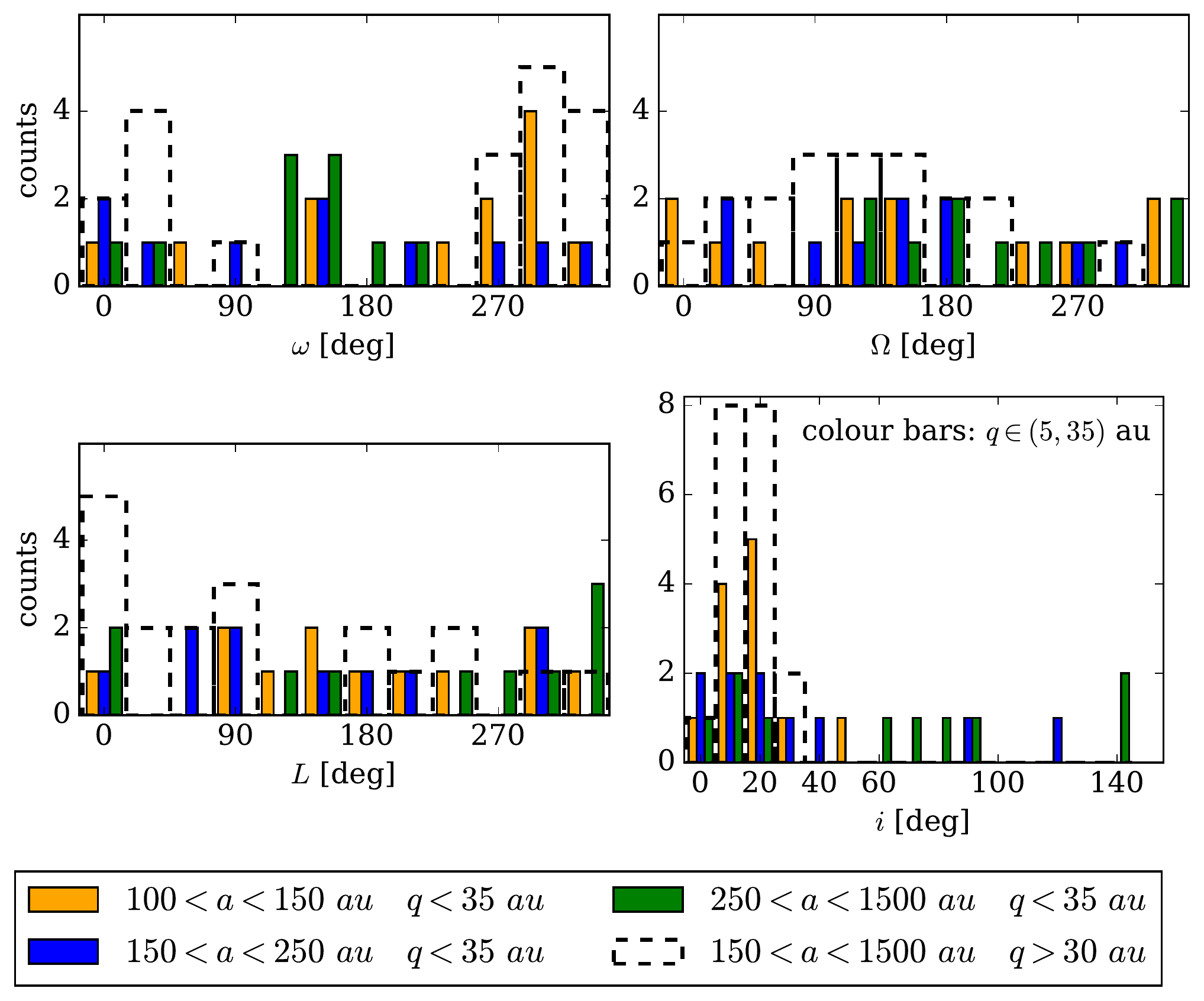}}
\caption{Histogram distributions of argument of perihelion (top left panel), longitude of ascending node (top right panel), perihelion longitude (bottom left panel), and inclination (bottom right panel) inferred from the well-determined orbits of TNOS and Centaurs fulfilling the criteria $100<a<1500$ au, $q<35$ au (colour histograms). The distributions inferred from all well-determined orbits of ETNOs ($150<a<1500$ au, $q>30$ au) are indicated as dashed bars.}
\label{img:data2}
\end{center}
\end{figure}

Here we discuss clues to the existence of the external perturbation, which give us clues to the existence of the external perturbation.
\cite{ST16} noted that the value $q=35$ au divides visually the observed population of objects with $a>150$ au and perihelia among and beyond giant planets in two distinct populations. 
This $q$-filter removes five objects (2013 UH$_{15}$, 2010 VZ$_{98}$, 2001 FP$_{185}$, 2013 FS$_{28}$, and 2015 SO$_{20}$) from the ETNOs sample. One of these objects, 2013 FS$_{28}$, is an outlier in terms of $\omega$, with $\omega=101.5$ deg \citep{ST16}.
Histogram distributions of the orbital elements $\omega$, $\Omega$, and $i$, as well as perihelion longitude $L$ for $100<a<1500$ au, $q>35$ au TNOs listed in the MPC orbit database on September 1, 2016 (colour histograms) are depicted in Fig. \ref{img:data1}.
The largest $a$ of all ETNOs has Sedna, $a\approx500$ au, i.e. no TNOs with $500<a<1500$ au, $q>30$ au are known.
Figure \ref{img:data2} considers the latter group of $100<a<1500$ au, $q<35$ au orbits. These could be substantially influenced by Neptune and the other giant planets.
In order to illustrate the effect of setting the boundary $q$ to be $q=35$ au, we also show, in both figures, the distributions inferred from all well-determined orbits of ETNOs (dashed histogram); ETNOs were defined in \cite{TS14} as objects with $150<a<1500$ au, $q>30$ au. We have to bear in mind that this choice of the boundary $q$ may be just our means of forcing the data to behave as we expect.

The two $q-$populations are distinct in terms of $\omega$ and $\Omega$ when $a>150$ au. In the case of $\omega$, there are no ETNOs with $q>35$ au, but many objects with $a>150$ au, $q<35$ au, occupying the region around $\omega=180$ deg. Object 2003 SS$_{422}$ is listed in the NASA Jet Propulsion Laboratory (JPL) small body database with orbital elements (and their $1\sigma$ uncertainties) at Epoch 2457600.5 (July 31, 2016): $a=193.5\pm47.8$ au, $q=39.4\pm1.0$ au, $i=16.8\pm0.1$ deg, $\omega=209.8\pm17.2$ deg, and $\Omega=151.1\pm0.1$ deg. This object was not counted because of its large uncertainty in $a$. If this object would be qualified as ETNO with future observations, then it would be the first ETNO with $q>35$ au with $\omega$ near 180 deg, which is about 180 deg away from the presently established clustering position. Additionally, it seems that $q<35$ au objects prefer higher values of $\Omega$ than their $q>35$ au counterparts. There is no obvious distinction between the two populations in terms of $L$. The group of lesser perihelia also comprises objects on highly inclined and retrograde orbits.

Looking at TNOs in Fig. \ref{img:data1}, the non-uniformity of the distributions is apparent, with a non-uniformity degree being $a$-dependent, as can be expected when an external perturbation is present. Orbits with low inclinations (<10 deg) and $q>35$ au are very rare, regardless of $a$, which is in stark contrast with the expected observational bias (e.g. \citealp{dlFdlF14}).

On September 1, 2016, the MPC orbit database listed 40 TNOs with $a>100$ au, 19 ETNOs, and 8 long-periodic ETNOs with $a>250$ au. When we restrict ourselves to $q>35$ au TNOs only, these numbers reduce to 32 TNOs with $a>100$ au, 14 ETNOs, and the same number of 8 long-periodic ETNOs.
Clearly, we are dealing with small-number statistics.
However, as pointed out elsewhere \citep{BF16,BB16}, it is unlikely that the observed clustering is purely due to chance.
Assuming the $\omega$, $\Omega$, and $L$ of ETNOs are uniformly distributed, the circulation of these angles is ensured by the perturbations from the giant planets, the probability of the observed clustering in $\omega$, $\Omega$, and $L$, beyond $a=250$ au level, is $0.1\%$, $3.0\%$, and $5.0\%$, respectively.
These estimates come from a following Monte Carlo simulation a la \cite{BB16}: we drew 8 angles, $\theta_{n}$ , which we call a set, randomly from a uniform distribution in range from 0 to 360 deg, $10^{6}$ times, and calculated the circular variance of each set, $\sigma^{2}_{i}=1-R$, $R=\left(\overline{S}^{2}+\overline{C}^{2}\right)^{1/2}$, $\overline{S}=\left(1/8\right)\sum_{n=1}^{8}\sin\theta_{n}$, $\overline{C}=\left(1/8\right)\sum_{n=1}^{8}\cos\theta_{n}$.
The probability is given as a frequency of sets for which $\sigma^{2}_{i}$ is smaller than the circular variance of the observational data.

\cite{BF16} plotted the distribution of $\omega$ for the whole known TNOs population (taking the data from the JPL\ small body database) and for a subset of TNOs that might not be in mean motion resonances with Neptune.
These authors found that the distribution of $\omega$ is well modelled, especially in the latter case, by overlapping normal distributions, peaking at 0 (360) and 180 deg, with standard deviations of 60 deg. This could be the effect of observational bias \citep{dlFdlF14,ST16}. Using the mentioned distribution made of the overlapping normal distributions, the Monte Carlo estimate of the probability of the observed clustering in $\omega$ is $0.5\%$ for $a>250$ au orbits.

Observational bias can be an important aspect of the puzzle.
A thorough examination of biases and discussion on this topic can be found, for example in \cite{ST16}. The known bias in $\omega$ is twofold: (1) if one does observations close to the ecliptic (as many surveys do), $\omega$s close to $\omega=0$ (360) deg and 180 deg are preferred because then the perihelion of an object, which is the most likely position in which an object is discovered, is close to the ecliptic and (2) observations from the southern hemisphere tend to prefer $\omega\in(180,360)$ deg, and by contrast, $\omega\in(0,180)$ is preferred when observations are made from the northern hemisphere. By employing off-ecliptic observations, made primarily from Chile, and taking into account the fact that there are no $q>35$ au ETNOs with $\omega$ close to 180 deg, \cite{ST16} concluded that the clustering in $\omega$ for $q>35$ au ETNOs is real with high probability. 

The parameter $L$ can be biased as well. New objects are most likely to be discovered near perihelion, and most surveys avoid the star polluted Galactic plane. Hence, there are two main time windows of the year to carry out a survey\footnote{These are spring and autumn preferentially. By inspecting dates of ETNOs discoveries, one finds that they were discovered primarily in spring and autumn.}, which possibly leads to bias in $L$. This bias should suppress objects near the ecliptic with right ascension, and hence $L$, around 90 deg. This is where all ETNOs with $a>250$ au, $q>35$ au, but one are found; there is another one found around 270 deg, see Fig. 3 in \cite{ST16}. 
If the observed population is biased in this way in $L$, then the external perturbation scenario is further promoted.
With $\omega$ clustered around $\omega\approx340$ deg and $L$ biased, $\Omega$ is biased as well. The Galactic plane avoidance then suppresses ETNOs with $\Omega$ around 110, where oddly\footnote{This is only odd if there is no substantial external perturbation present.} the majority of ETNOs reside, and around 290 deg. The asymmetry of the $\Omega$ distribution cannot be explained by the intrinsic declination bias, as this is symmetric with respect to the preferred value $\Omega=180$ deg \citep{dlFdlF14}.

\begin{figure}
\begin{center}
\resizebox{\hsize}{!}{\includegraphics{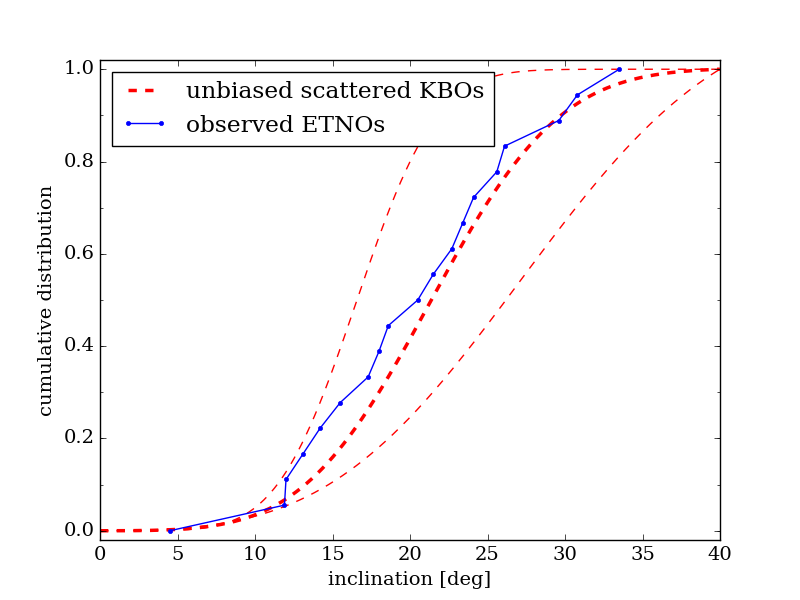}}
\caption{Unbiased cumulative distribution function of the scattered KBOs (dashed line) vs. observed cumulative distribution of ETNOs (solid line), both as a function of inclination. The thick dashed line represents the best-fit model of \cite{Gul+10}, while the thinner dashed lines are its 1$\sigma$ variations. The dots correspond to the observed inclinations of ETNOs and are equidistantly spaced between 0 and 1 on the vertical axis.}
\label{img:bias_i}
\end{center}
\end{figure}

Observational bias prefers low inclinations \citep{dlFdlF14}. 
Hence, the clustering in $i$ is inconsistent with being an observational effect. Interestingly, the (biased) inclination distribution of the observed ETNOs resembles the unbiased inclination distribution of the population of scattered Kuiper belt objects (KBOs), high $a$ and $e$ KBOs\footnote{ETNOs are a subset of the scattered KBOs.}. We illustrate this in Fig. \ref{img:bias_i}, where we show an unbiased cumulative distribution function of the scattered KBOs (dashed line) versus observed cumulative distribution of ETNOs (solid line) as a function of inclination. For scattered KBOs, we used the model from \cite{Gul+10}, modelling the inclination distribution as a Gaussian multiplied by $\sin i$ \citep{Bro01}, centred at $19.1\substack{+3.9 \\ -3.6}$ deg, and with a width of $6.9\substack{+4.1 \\ -2.7}$ deg. Cumulative distribution for ETNOs was constructed in a simplified manner. The dots in Fig. \ref{img:bias_i} correspond to the observed inclinations of ETNOs and are equidistantly spaced between 0 and 1 on the vertical axis. We conclude that ETNOs inclination distribution is probably not significantly biased. But still, we have to bear in mind the preference of low inclinations and that the width of the real ETNOs inclination distribution is somewhat higher than that inferred from Fig. \ref{img:bias_i}.

In what follows, we consider statistics about the orbital elements  $\omega$, $\Omega$, and $i$ of the ETNOs as not significantly biased and representative of the real (unbiased) population.

\cite{BB16} and also \cite{ST16} suggested taking into account only ETNOs that are dynamically stable for the age of the solar system. 
This is a good idea in general.
However, it is misleading to investigate the stability of ETNOs in a dynamical model with giant planets alone and no external perturbation, as carried out by \cite{ST16} and \cite{BB16}, and then infer properties of the external perturbation from such a ``stable'' sample. We do not distinguish between stable and unstable ETNOs in this way, but we do distinguish them on $q>35$ au and $q<35$ au ETNOs.

\section{Milgromian dynamics}\label{sec:MOND}

In 1983 Milgrom proposed a simple modification of the standard dynamics attributed to low accelerations \citep{Mil83b}.
He did so to explain dynamical discrepancies in the Universe manifested at that time mainly through asymptotically flat rotation curves \citep{Bos81,R+82}.
He assumed acceleration to be calculated not from ${\bf g} = {\bf g^{N}}$ but
from\begin{eqnarray}\label{MOND_basic}
{\bf g} = \nu( g^{N}/a_{0}){\bf g}^{N}~,
\end{eqnarray}
where ${\bf g}^{N}$ is expected Newtonian gravitational acceleration, $g^{N}$ is its norm, $\nu(y)$, $y\equiv g^{N}/a_{0}$ is a modification factor, the so-called interpolating function, fulfilling behaviour $\nu(y)\rightarrow 1$ for $y\gg1$ (Newtonian limit) and $\nu(y)\rightarrow y^{-1/2}$ for $y\ll1$ (deep-MD limit), and $a_{0}\sim 10^{-10}$ m s$^{-2}$ is a new constant of nature.
He realised that such modification leads to a whole new phenomenology, so that low acceleration 
systems, such as galaxies, should display some special, systematic properties \citep{Mil83a,Mil83b}. At that time, when the amount of data on galaxies, and especially on low surface brightness (LSB) galaxies\footnote{Systems in deep-MD regime when isolated.}, was limited, only one observation, which is actually a premise of writing down Eq. (\ref{MOND_basic})  -- asymptotically flat rotation curves -- was recognised as one of these special properties. Today, when various types of galaxies, including LSB galaxies, were discovered and scrutinised, we now know that many Milgrom's predictions from his 1983 papers were fulfilled (reviewed e.g. in \citealp{FM12}).

Milgrom and others have already developed generally covariant modified gravity theories, where Eq. (\ref{MOND_basic}) is implemented as a static weak field limit in the case of systems with spherically symmetrically distributed mass. In these theories the weak field regime is governed by a modified Poisson equation. Two types of the modified Poisson approach are common. In the first type, known as aquadratic Lagrangian theory (AQUAL), the modified Poisson equation is written \citep{BM84}
\begin{eqnarray}\label{AQUAL}
\nabla\cdot\left[\mu\left(\frac{\vert\nabla\Phi\vert}{a_{0}}\right)\nabla\Phi\right]~=~4\pi G\rho~=~\nabla^{2}\Phi_{N}~,
\end{eqnarray}
where the latter equality is the classical Poisson equation with Newtonian potential $\Phi_{N}$, (baryonic) matter density $\rho$, and Newton's gravitational constant $G$, $\mu$ is inverse of $\nu$ from Eq. (\ref{MOND_basic}), and the equation of motion is ${\bf g}=-\nabla\Phi$. The second modified Poisson approach, known as quasi-linear MOND (QUMOND), states that the governing gravitational potential $\Phi$ is given by \citep{Mil10}
\begin{eqnarray}\label{QUMOND}
\nabla^{2}\Phi~=~\nabla\cdot\left[\nu\left(\frac{\vert\nabla\Phi_{N}\vert}{a_{0}}\right)\nabla\Phi_{N}\right]~.
\end{eqnarray}
Having $\mu$ as inverse of $\nu$, the two theories coincide in the spherically symmetric case, giving rise to Eq. (\ref{MOND_basic}) \citep{Mil10}. Outside of the spherical symmetry the two theories are not equivalent.

Eq. (\ref{MOND_basic}) can equivalently come from modified inertia instead of the modified gravity discussed above,
 where modified is the kinetic portion of the classical action, hence
the equation of motion. \cite{Mil94} showed that such theories always have to be time nonlocal to be Galilean invariant. As of this complication there is no full-fledged modified inertia theory developed yet (but see discussion in section 7.10 of \citealp{FM12}). In modified inertia approach, Eq. (\ref{MOND_basic}) holds exactly in the case of circular orbits alone.

\subsection{Effect of Milgromian dynamics in the solar system}\label{sec:MDinSS}

Three effects of MD are recognised to act in the solar system. The three effects are separable as long as $r\ll r_{M}$, where $r_{M}\equiv(GM_{\odot}/a_{0})^{1/2}$ \citep{Mil09,Mil12}.
\begin{itemize}
\item Enhanced gravity effect
\end{itemize}
This classical MD effect results from departure of the MD interpolating function from unity. In the case of a spherical system, the anomalous acceleration due to this effect, $\delta{\bf g} = {\bf g} - {\bf g}^{N}$, can be written as
\begin{eqnarray}\label{mond_effect1}
\delta {\bf g} = \left[\nu(g^{N}/a_{0}) - 1\right]{\bf g}^{N}~.
\end{eqnarray}
This effect successfully encapsulates various aspects of Galactic phenomenology \citep{Mil83b,FM12}. This effect strongly depends on the choice of the interpolating function and value of $a_{0}$, and can be made arbitrarily small by a suitable choice of the interpolating function.
\begin{itemize}
\item External field effect (EFE)
\end{itemize}
This effect is inherent to modified gravity formulations of MD\footnote{This effect does not appear in the modified inertia formulations of MD.} and it comes from the non-linearity of the modified Poisson equation. An embedding external field that is constant and uniform, with magnitude $g_{e}$, $g_{e}\lesssim a_{0}$, influences internal dynamics in an embedded system (for more details see e.g. \citealp{PK16,FM12}).
Contrary to the enhanced gravity effect, EFE appears even in the case of high internal accelerations ($\mu$ and $\nu$ arbitrarily close to 1). 
The solar system is falling freely in the external gravitational field of the Galaxy. The external field has magnitude $g_{e}\approx v^{2}_{0}/R_{0}\approx 240^{2}$ km$^{2}$ s$^{-2}$ $/~8.0$ kpc $\doteq 2.0$ $\times 10^{-10}$ m s$^{-2}$ (for values of the Galactic parameters $v_{0}$ and $R_{0}$, we refer to \citealp{McM+10, Sch12}). Close to the Sun, at distances $r\ll r_{M}$, $r_{M}\equiv(GM_{\odot}/a_{0})^{1/2}\sim 7000$ au, where $r_{M}$ is characteristic radius in MD solar system, EFE expresses itself dominantly as an anomalous quadrupole, $\delta\Phi$, so a governing potential entering the equation of motion is written\begin{eqnarray}\label{mond_effect2}
\Phi &=& \Phi_{N} - \frac{Q_{2}}{2} r^{i}r^{j}\left(e_{i}e_{j}-\frac{1}{3}\delta_{ij}\right)\nonumber \\
     &=& \Phi_{N} + \delta\Phi~,
\end{eqnarray}
where ${\bf e}\equiv {\bf g}_{e}/g_{e}$ is a unitary vector pointing towards the Galactic centre\footnote{In AQUAL, ${\bf e}$ is pointing in the direction of the real (Milgromian) gravitational field of the Galaxy. In QUMOND, ${\bf e}$ is pointing in the direction of the Newtonian gravitational field. For the approximately axially symmetric Galaxy, these directions are nearly the same.} (GC) and $\Phi_{N}$ is the expected Newtonian potential \citep{Mil09,BN11}. The strength of the anomaly at a given position is represented by the magnitude of the parameter $Q_{2}$, which depends on the specific modified Poisson formulation, form of the interpolating function, and strength of the external field in natural units $a_{0}$ \citep{Mil09,BN11}. If not stated otherwise, $Q_{2}$ is assumed to be positive. All routinely used forms of the interpolating function yield\footnote{In \cite{Mil09}, a dimensionless parameter $q$, $q=-2Q_{2}(GM_{\odot})^{1/2}/(3a^{3/2}_{0})$, was employed instead of $Q_{2}$.} $Q_{2}>0$ \citep{Mil09,BN11}. If $Q_{2}$ is taken to be a constant
in some region of interest, the value $\delta\Phi$ has the same functional form  as the tidal potential coming from the hypothetical P9 or that from the Galaxy.
We do not know the exact form of the interpolating function, therefore we do not know the value of $Q_{2}$, but to be dynamically important in the ETNOs region it would have to be much stronger than the quadrupole coming from the Galactic tides. 
If we consider values of $Q_{2}$ close to the the theoretical maximum allowed by the Earth-Cassini spacecraft range data (hereafter Cassini data), $Q_{2}\sim 10^{-27}$ s$^{-2}$ \citep{H+14}, then EFE is $\sim10^{3}$ times stronger than the Galactic tides \citep{H+14}. Instantaneous action of EFE is the same as that of P9 lying in the direction of the GC-anticentre, characterised by the tidal parameter $GM'/r'^{3}=Q_{2}/3$, where $M'$ is mass of P9 and $r'$, $r'\gg r$ is its heliocentric distance \citep{Ior10}. 
Thus, EFE and the dynamical effect of P9 are indistinguishable on short timescales in the ETNOs region. Intriguingly, perihelia of ETNOs with $a$ beyond 250 au cluster close to the direction of the present-day Galactic anticentre (the preferred direction in Eq. (\ref{mond_effect2})). Comparison between orbital characteristics of ETNOs induced on long timescales by EFE and those induced by P9 is a topic of this study.

\begin{itemize}
\item Asphericity effect
\end{itemize}
This effect appears only in modified gravity formulations of MD. It affects every system with non-spherically symmetric matter distribution. Suppose we have an isolated high-acceleration system S, which assumes $\mu=\nu=1$ to the desired accuracy everywhere within S, of total mass $M$ and extension $R$, $R\ll r_{M}$, where $r_{M}=\sqrt{GM/a_{0}}$. Suppose also that mass distribution within S, $\rho({\bf r})$, varies on timescales much larger than $r_{M}/c$, where $c$ is the speed of light. The classical Poisson equation coincides with MD equations in S. But, the Poisson equation is not valid everywhere up to infinity. It is definitely not valid for $r> r_{M}$. Hence, the solution of the Poisson equation in S is slightly different from the Newtonian solution.
In QUMOND, \cite{Mil12} showed that dynamics in S is governed by the potential
\begin{eqnarray}\label{mond_effect3}
\Phi   &=& \Phi_{N} - \frac{\alpha G}{r^{5}_{M}} r^{i}r^{j}Q_{ij}\nonumber \\
       &=& \Phi_{N} + \delta\Phi~,\nonumber \\
Q_{ij} &=& \frac{1}{2}\int\rho({\bf r'})\left(r'^{2}\delta_{ij}-3r'_{i}r'_{j}\right)d^{3}r'~,
\end{eqnarray}
where $\Phi_{N}$ is the expected Newtonian potential, $\delta\Phi$ is a dominant MD correction from the asphericity effect, a quadrupole field, and $\alpha$, typically $\alpha\sim 1$, is a numerical factor that depends on the form of the interpolating function. In the case of the solar system\footnote{Solar system is not an isolated system, so we have a superposition of EFE and the asphericity effect.},  $M\approx M_{\odot}$ and $Q_{ij}$ is the quadrupole moment of the solar system.

In the standard formulation of MD and in many parent relativistic theories of MD, this effect is too weak to be detected in the solar system \citep{Mil12}. In special theories where Newtonian regime is not reached beyond $g^{N}\sim a_{0}$, but beyond $g^{N}\sim \kappa a_{0}$, $\alpha$ in Eq. (\ref{mond_effect3}) becomes enhanced by $\kappa^{2}$ approximately \citep{Mil12}. An example of such a theory is, for instance TeVeS \citep{Bek04}, where $\kappa$ must have very high value to become compatible with general relativity \citep{Mil12}. However, the existence of two different acceleration constants in the theory, $a_{0}$ and $a^{*}_{0}\equiv\kappa a_{0}$, seems rather unnatural.
The anomalous perihelion precession of Saturn constrains $\kappa$ to be $\kappa\leq 3500$ \citep{Ior13}.

\section{Benchmark model}\label{sec:model}

We start with construction of a simple yet realistic dynamical model of the solar system in MD. We call this model a benchmark model. The region of our interest is a sphere of radius 2000 au that is centred on the Sun. This region is large enough to investigate the orbit of ETNO in its entirety and also small enough for Eq. (\ref{mond_effect2}) to be applicable;
Sedna,
a body with the largest aphelion distance, cruises up to $\sim900$ au.
 
By combining an analysis of rotation curves of galaxies and solar system constraints\footnote{The solar system constraints in \cite{H+16} come from EFE quadrupole anomaly, all the other MD effects are negligible in the planetary zone \citep{H+14,H+16}.} provided by the Cassini spacecraft, \cite{H+16} showed that many popular MD interpolating functions are not allowed by the data. These authors also listed examples of the allowed interpolating functions (their Table 2). Generally speaking, these allowed functions are characterised by extremely rapid transition into the Newtonian regime.\footnote{Also, these are the functions for which the enhanced gravity effect is negligible in the planetary region, compared to EFE.} Referring to Table 2 in \cite{H+16}, we chose $\overline\nu_{n}(y)=\left[1-\exp(-y^{n})\right]^{-1/(2n)}+\left[1-1/(2n)\right]\exp(-y^{n})$ interpolating function family as  most representative. The Cassini data restrict $\overline\nu_{n}$ to $n\geq 2$. The enhanced gravity effect is completely negligible in the whole solar system\footnote{Except for small regions, where internal and external gravitational fields add up into a vector of small magnitude.} for $\overline\nu_{n}$, $n\gtrsim 3$.
Thus, in the first approximation, we neglect the enhanced gravity effect in the benchmark model.
 
The parameter $Q_{2}$ of Eq. (\ref{mond_effect2}) is position dependent in general \citep{Mil09}. However, as shown by \cite{BN11} in the framework of AQUAL, $Q_{2}$ can be considered a constant in a large sphere centring on the Sun. The variation of $Q_{2}$ within our volume of interest is up to few percent \citep{BN11}. 

In the benchmark model, the only MD effect we take into account is EFE quadrupole anomaly. We model EFE as in Eq. (\ref{mond_effect2}), with position independent $Q_{2}$, $Q_{2}=Q_{2}(0)$. All higher multipoles are omitted. The value of $Q_{2}$ was constrained by the Cassini data to be $Q_{2}=3\pm 3\times 10^{-27}$ s$^{-2}$ (nominal $\pm 1\sigma$ value) \citep{H+14}. From now on, if we refer to $Q_{2}$ without explicitly stating its units, the unit is $10^{-27}$ s$^{-2}$.
The unit vector ${\bf e}$ of Eq. (\ref{mond_effect2}), pointing towards the GC, is assumed to circulate uniformly  with time according to
\begin{eqnarray}\label{unit_vector}
e_{x-g} &=& +\cos(w t + \tau) \nonumber \\
e_{y-g} &=& -\sin(w t + \tau)\nonumber \\
e_{z-g} &=& 0 \nonumber \\
e_{x} &=& -~0.054876~e_{x-g} + 0.494109~e_{y-g}\nonumber \\
e_{y} &=& -~0.993821~e_{x-g} - 0.110991~e_{y-g}\nonumber \\
e_{z} &=& -~0.096477~e_{x-g} + 0.862286~e_{y-g}~,
\end{eqnarray}
where the first three lines are components of ${\bf e}$ in the Galactic coordinates\footnote{Principal axis points from the Sun to the GC.} and the last three lines represent transformation from the Galactic to the heliocentric ecliptic coordinates at J2000.
We assumed that the Sun is moving in the circular orbit around the GC\footnote{The Sun is moving in the clockwise direction, looking from the north Galactic pole perspective.}, lying in the Galactic mid-plane, with angular speed $w$; $\tau$ is the phase of the motion.

\section{Orbital clustering:\ Initial numerical exploration}\label{sec:model0}

Our aim is to investigate whether orbits of primordial ETNOs are driven to long-living orbital confinement in the solar system ruled by the dynamical laws of MD. 
For this purpose we made use of the benchmark model. Additionally, we model the secular effect of the giant planets by considering the Sun of radius $R=a_{U}$, where $a_{U}$ is the semimajor axis of Uranus and $J_{2}$ is the moment of magnitude $J_{2}=\sum_{i}m_{i}a^{2}_{i}/(2M R^{2}),$ and where we sum over the giant planets with masses $m_{i}$, semimajor axes $a_{i}$, and $M$ is mass of the Sun \citep{BB16}. Close encounters with the giant planets were completely omitted in this initial numerical exploration.

We traced the evolution of primordial ETNOs in $\omega-e$, $\Omega-e$, and $L-e$ plane, considering $Q_{2}$ taken from a set $Q_{2}=\{1.0,~2.0,~4.0,~6.0\}$. In MD, the external perturbation, substituting a distant planet, is the quadrupole anomaly arising from EFE; see Eqs. (\ref{mond_effect2}) and (\ref{unit_vector}). The direction of the perturbation is known and we alter only the parameter $Q_{2}$.

The ETNOs were modelled as test particles and followed for 4 Gyr. We consider three values of initial semimajor axis: $a=150$, 350, and 550 au. For each value of $a$, 20 test particle orbits were initialised with an $e$ spanning range (0, 0.95) with a constant increment of 0.05 \footnote{In the case $a=150$ au, the particles in $e=0.90$ and $e=0.95$ orbits were doomed as $q<a_{U}$ holds for them at the time of their initialisation.}, $i$ drawn randomly from a uniform distribution in range (0, 5) deg, $\omega$ drawn randomly from a uniform distribution in range (0, 360) deg, and $\Omega$ calculated so that $L=90$ deg. Similarly, another 20 particles with $L=270$ deg were set up for each value of $a$. The particles were started in perihelion. The values $L=90$ and 270 deg were identified as stable libration regions with trial and error method. We note that $L=270$ -- 90 deg is also roughly the direction of the present-day GC--anticentre. The ecliptic latitude of the GC is only minus few degrees. Here our intention is to demonstrate the existence of the long-living libration regions; in the next section we examine development of the orbital clustering, starting with completely randomised orbits.

Every orbital integration in this paper was carried out using \mercury N-body integration software package \citep{Cha99}. The hybrid symplectic-Bulirsch-Stoer algorithm with a timestep equal to a tenth of the orbital period of Neptune was employed. Phase $\tau$ in Eq. (\ref{unit_vector}) was chosen in a way that at $t=4$ Gyr the vector ${\bf e}$ points towards the present-day GC. We tested that the value of $\tau$ actually had no influence on the results. 
If a particle had reached the distance 2000 au from the Sun, it was immediately discarded from the simulation\footnote{In fact, none of the particles experienced such large change in $a$. The discarded particles typically came down to $r=a_{U}$.}. 
There is no reason to discard particles beyond 2000 au in the Newtonian simulation, but we wanted to apply the same measure in both frameworks. If a particle had come closer than $r=a_{U}$, it was discarded from the simulation as well.

For the sake of comparison, we ran also two P9 simulations in Newtonian dynamics. In the first simulation, P9 of mass $M'=10~M_{\oplus}$ was assumed to orbit in $a'=700$ au, $e'=0.6$, $i'=0$ deg, and $L'=240$ deg orbit \citep{BB16}. The primed quantities always refer to P9.
The test particle orbits were initialised with $L-L'$ equal to 0 (20 particles for each $a$) and 180 deg (20 particles for each $a$), and nil inclinations\footnote{In this case, it is meaningful to examine evolution in the $L-e$ plane only, as all inclinations were set to zero.}. All the other orbital elements were set up as in the MD simulation. P9 was initialised in aphelion. We also did a simulation assuming $i'=30$ deg and inclined test particle orbits with $i$ drawn randomly from a uniform distribution in range (0, 5) deg. In addition to the previous rules for discarding particles, we also discarded all particles that come within one Hill radius from P9.

\begin{figure*}
\begin{center}
\resizebox{\hsize}{!}{\includegraphics{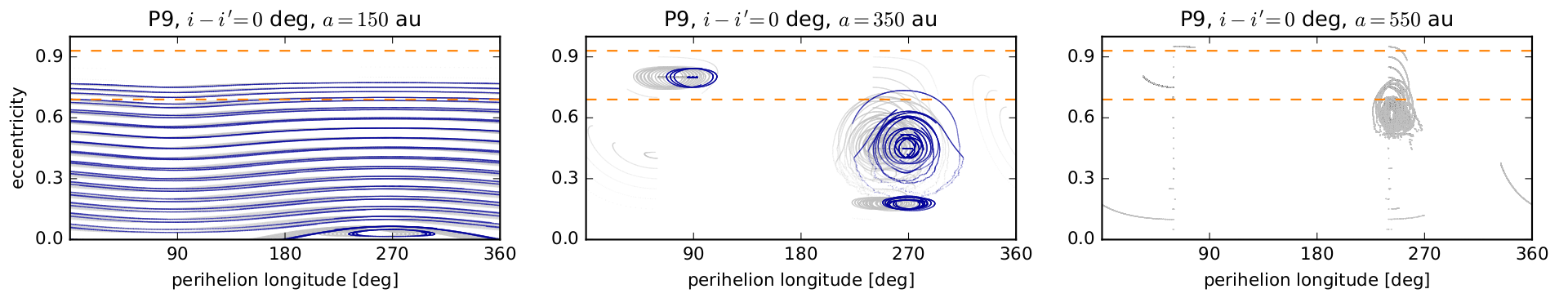}}
\caption{Evolution of test particles in $L-e$ plane as inferred from the P9 simulation. The particles and P9 orbit the Sun in the same plane by design. The eccentricity range of the observed ETNOs is indicated by the dashed horizontal lines. Two transparency levels are used as a proxy for two different time windows of the simulation. The less transparent evolutionary tracks represent long-living particles in the last Gyr of the simulation. The more transparent tracks refer to evolution in the first 3 Gyr of the simulation and also comprise unstable orbits. Three starting semimajor axes were used as follows: $a=150$ au (first column), $a=350$ au (second column), and $a=550$ au (third column). Nbody encounters are completely omitted in this case, hence semimajor axes do not change during the simulation. The secular effect of the giant planets is modelled with a strong $J_{2}$ moment of the Sun. There were no stable orbits at $a=550$ au level.}
\label{img:mcp9L}
\end{center}
\end{figure*}

We depict evolution of test particles in $L-e$ plane as inferred from the first P9 simulation in Fig. \ref{img:mcp9L}. All particles meet the condition $i-i'=0$ by design. The discussed orbital elements always refer to the heliocentric ecliptic J2000 reference frame. Two transparency levels are used as a proxy for two different time windows of the simulation. The less transparent evolutionary tracks represent long-living particles in the last Gyr of the simulation. The more transparent tracks refer to evolution in the first 3 Gyr of the simulation and also comprise unstable orbits. Orbits with $a=150$ au typically experience trivial apsidal circulation. Only one small libration island exists at $L\approx240$ deg (aligned with P9) for near-circular orbits. For $a=350$ au, two long-term stable libration regimes exist \citep{BB16}: first, the resonant dynamics regime concerning eccentric orbits (with eccentricities similar to those of the observed ETNOs) and centred at $L\approx60$ deg and anti-aligned with P9 and, second, the secular dynamics regime concerning less eccentric orbits, centring at $L\approx240$ deg and aligned with P9. These absolute positions come from our choice of $L'$, which was tuned to match the observed clustering position; see Fig. \ref{img:data1}. \cite{BB16} always refer to the relative longitude of perihelion $\varpi-\varpi'$ in their analysis, as they did not know the orbital elements of the hypothesised P9 a priori. Here, we build on the results of their pioneering numerical experiments. The shift of the libration centres in $L-e$ plane with time follows from the shift of $L'$ caused by the giant planets. At 550 au level, no stable libration develops.
In the second P9 test case, where we consider an inclined perturber with $i'=30$ deg and inclined test particle orbits, there were no long-term stable orbits.

The picture in MD is rich as well.
Orbits with $a=150$ au are characterised by trivial circulation of the angles $(L,~\Omega,~\omega)$. No stable libration develops at this $a$-level for realistically high value of $Q_{2}$. The values $a=350$ and 550 au are more interesting.
The orbits of particles are strongly confined in longitude of ascending node $\Omega$ in all 4 MD simulations. Both eccentric (ETNOs) and less eccentric particles, with both $a=350$ and 550 au, strongly prefer $\Omega\in(180,360)$ deg. In ETNOs eccentricity range (EER), $e\in(0.69, 0.93)$, stable, and often tight, libration around $\Omega\approx270$ deg occur, for instance where $Q_{2}=1$ or 2, $a=350$ au. 

Concerning perihelion longitude $L$, for $a=350$ au, we get a circulation of the angles in EER irrespective of $Q_{2}$. In the case of lower eccentricities, we get a circulation for models with $Q_{2}\lesssim4$, and neither circulation nor strong confinement using models with $Q_{2}\gtrsim4$. The specialty of values $L=90$ and 270 deg is still noticeable. 

\begin{figure*}
\begin{center}
\resizebox{\hsize}{!}{\includegraphics{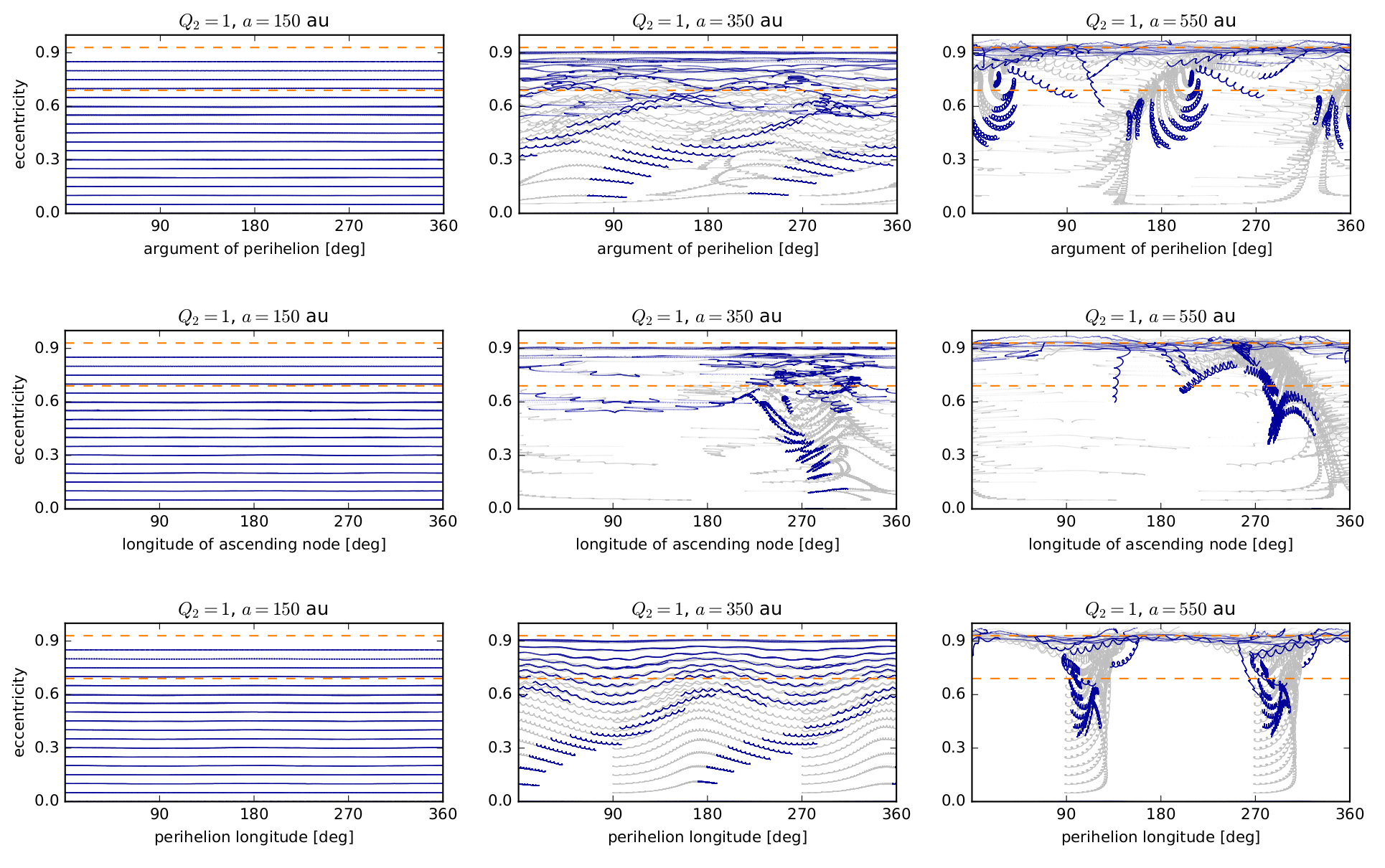}}
\caption{Evolution of test particles in $\omega-e$ (first row), $\Omega-e$ (second row), and $L-e$ (third row) plane as inferred from the benchmark model simulation with $Q_{2}=1$.}
\label{img:mc1}
\end{center}
\end{figure*}

The case $Q_{2}=1$, $a=550$ au is especially interesting. The motion in $L-e$ plane is confined to narrow regions near $L\approx90$ and $L\approx270$ deg in a broad range of eccentricities, spanning from low eccentricities to EER. For eccentricities $e\gtrsim0.9$, the circulation of the angles sets in. In $\omega-e$ plane, there are orbits clustered near $\omega\approx0$ and $\omega\approx180$ deg, below and at the low eccentricity end of EER.
We depict the evolution of test particles in $\omega-e$, $\Omega-e$, and $L-e$ plane as inferred from the benchmark model simulations using $Q_{2}=1$ in Fig. \ref{img:mc1}.

Despite the rich structure in $\omega-e$, $\Omega-e$, and $L-e$ planes in the MD, the simulated evolutionary paths are in sharp contrast with present-day observations. The simulations predict that the vast majority of ETNOs should have $\Omega\in(180,360)$ deg. Hence, the vast majority of ETNOs should have ascending node lying in the same half-space, with the GC lying in the same half-space. This is not an observed trend. Of 19\ ETNOs, 14 have $\Omega\in(0,180)$ deg, while 3 of them have $\Omega$ very close to 180 deg \citep{ST16}. 
Moreover, the tightness of the observed clustering in $\omega$ is not reproduced in our simple MD models. The discrimination of the orbits of particles according to their $q$ is of no meaning here as the giant planets were omitted in these initial simulations. 
In order to compare MD models with Newtonian models including P9, we should switch to a more realistic approach, also taking the gravitational interaction between ETNOs and Neptune into account.

\subsection{The case of negative $Q_{2}$}

The fact that the MD models predict the clustering region that lies roughly 180 deg ahead of the observed clustering in $\Omega$ motivates us to question the direction of the perturbation coming from EFE (effectively the sign of $Q_{2}$). 
When $Q_{2}$ is positive, a test particle is repelled from the Sun in the direction of the external field, which means that effectively the gravity of the Sun is reduced in that direction, while the particle is attracted to the Sun in the direction perpendicular to the external field, i.e. effectively the gravity of the Sun is enhanced in that direction.
In \cite{Mil09}, the author studied the effect MD has in the planetary region of the solar system. In the last section, he states that\footnote{Milgrom used $q$, $q\equiv-2Q_{2}(GM_{\odot})^{1/2}/(3a^{3/2}_{0})$, instead of $Q_{2}$, to quantify the strength of EFE.} although $Q_{2}$ was positive for all interpolating functions $\mu$ he studied, he was ``not able to prove that this is always the case from basic properties of $\mu$'' \citep{Mil09}. \cite{H+14} showed that the Cassini data yield $Q_{2}=3\pm3$ (nominal $\pm 1\sigma$ value).

We substituted $Q_{2}\rightarrow -Q_{2}$ and repeated the benchmark model simulations. This means we considered $Q_{2}$s that are 1 - 3$\sigma$ off of the nominal value of \cite{H+14}, with corresponding probabilities for $Q_{2}$ ranging from 16 ($Q_{2}<0$) down to 0.1$\%$ ($Q_{2}<-6$).

\begin{figure*}
\begin{center}
\resizebox{\hsize}{!}{\includegraphics{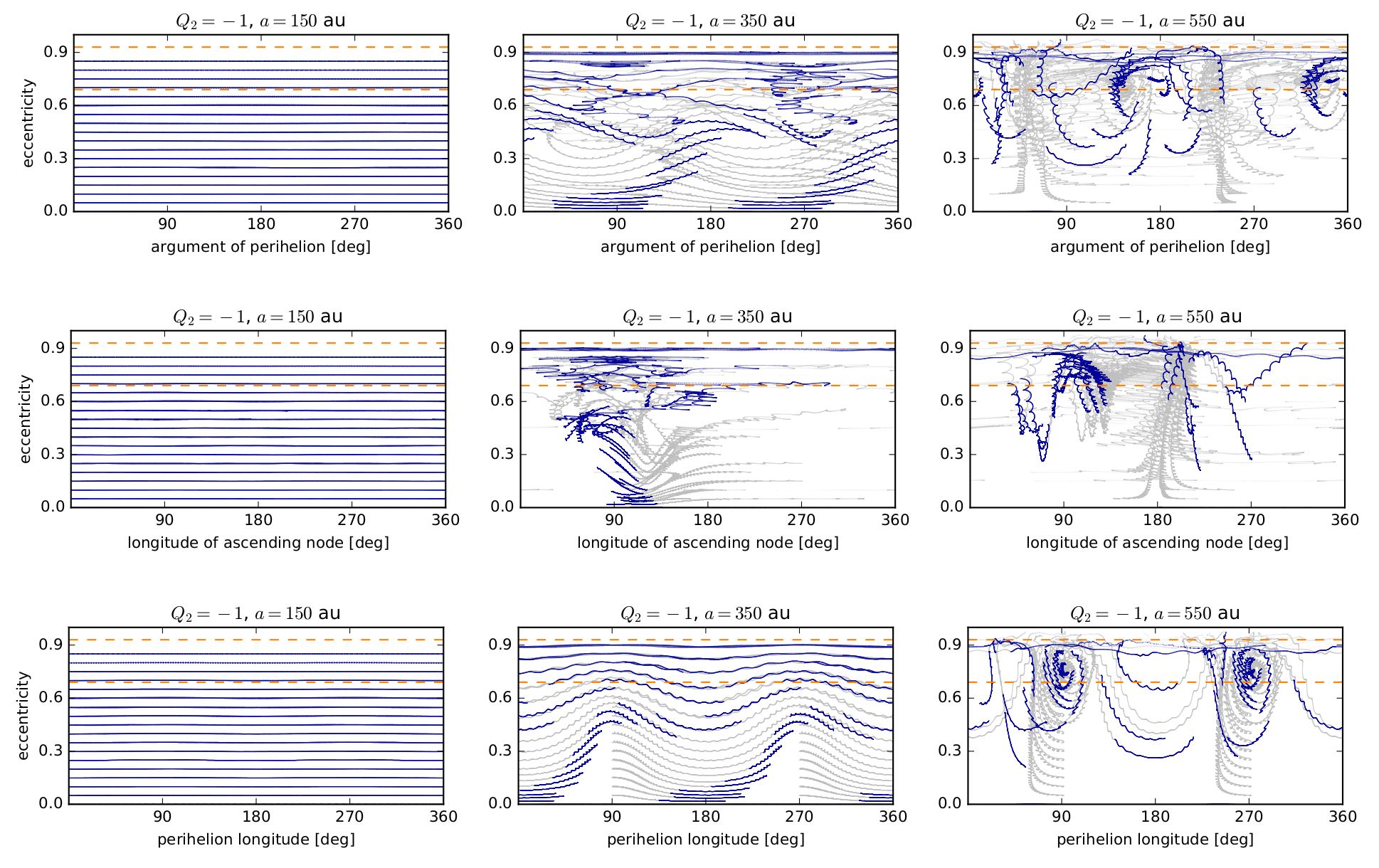}}
\caption{Evolution of test particles in $\omega-e$ (first row), $\Omega-e$ (second row), and $L-e$ (third row) plane as inferred from the benchmark model simulation with negative $Q_{2}$, $Q_{2}=-1$.}
\label{img:mcm1}
\end{center}
\end{figure*}

At $a=350$ au level, the substitution causes a shift of the clustering region by roughly 180 deg in $\Omega$, while leaving evolution in $\omega-e$ and $L-e$ planes without qualitative change. This can be seen in the case $Q_{2}=-1$ in Fig. \ref{img:mcm1}. 
Looking at EER at $a=550$ au, $\Omega$ lies sometimes preferentially in range $(0,180)$ deg ($Q_{2}=-1$), sometimes in range $(180,360)$ deg ($Q_{2}=-4$ and -6), and sometimes there is preference for neither of the two regions ($Q_{2}=-2$).

\section{Add Neptune and stir}\label{sec:results}

Neptune may have influenced the distribution of the orbital elements of ETNOs through orbital filtering (e.g. by preferentially ejecting  ETNOs in specific orbits) and resonant effects.
As a next step, we added Neptune orbiting the Sun to the dynamical model.
We consider the gravitation of the Sun and planet Neptune in the usual Newtonian way as a direct N-body interaction. 
The secular effect of the three remaining giant planets is modelled by considering the Sun of radius $R=a_{U}$, where $a_{U}$ is the semimajor axis of Uranus, and $J_{2}$ moment of magnitude $J_{2}=\sum_{i}m_{i}a^{2}_{i}/(2M R^{2})$, where we sum over the remaining giant planets with masses $m_{i}$, semimajor axes $a_{i}$, and $M$ is mass of the Sun \citep{BB16}.

We performed several numerical simulations, aiming to compare the dynamical effects of P9 and MD on orbits of primordial TNOs. 
Each simulation started with a thin disk of 400 test particles, where 40 particles had semimajor axis $a=100$ au, 40 particles $a=150$ au, and so on up to 40 particles with $a=550$ au. Perihelion distances of the particles with a given $a$ covered range $(30,50)$ au with a constant increment. The inclination, argument of perihelion, and longitude of ascending node of the test particle orbits were randomly drawn from a uniform distribution in range $(0,5)$ deg $(i)$, and  $(0,360)$ deg $(\omega, \Omega)$. The particles were initialised at their perihelia.
We also ran all the simulations with twice as many particles with an additional 400 particles initialised by the same procedure as delineated above to test the effect of the test particle number.

In the benchmark model simulations, we considered various values of $Q_{2}$ taken from a set $Q_{2}=\{0.2,~0.5,~1.0,~2.0,~4.0,~6.0\}$. For the sake of comparison, simulations with P9 in Newtonian dynamics (no EFE $\Leftrightarrow$ $Q_{2}=0$) were carried out. Two initial P9 orbits were considered: (\#1) $a'=700$ au, $e'=0.6$, $i'=30$ deg, $\omega'=150$ deg, $\Omega'=113$ deg \citep{BB16, BB16b, Fie+16}; (\#2) the same\footnote{Starting values of $a'$, $e'$, and $i'$ were altered only negligibly during the time of the simulation.} $a'$, $e'$, $i'$ as for orbit (\#1), but $\omega'=100$ deg, $\Omega'=140$ deg. The orbit (\#2) orbital elements account for variation in $\omega'$ and $\Omega'$, caused by the giant planets, and at $t=4$ Gyr the final $\omega'$ and $\Omega'$ were very close to the orbit (\#1) starting values.

\subsection{Sednoids}

The discovery of Sedna \citep{Bro+04} and 2012 VP$_{113}$ \citep{TS14} was surprising as their orbits did not fit into the standard picture of the solar system, settled in the current Galactic environment, containing a set of known planetary bodies. 
The literature discussed scenarios in which these orbits were shaped in environments that, in the past, were different from the present-day environment, for instance the possibility that the Sun was born in a star cluster \citep{KQ08,Bra+12} and/or was traveling much closer to the GC in the past \citep{Kai+11}. 

Another possibility is that the torque necessary to elevate perihelia of these bodies comes from an unknown planet-mass body, or maybe multiple bodies, beyond Neptune \citep{Gom+06,BB16}. 
In MD, similar orbits could be induced by EFE with no need for an additional external body \citep{PK16}. Both options have in common that they operate continuously. Hence, they produce a ``live'' population in which perihelia are repeatedly, and slowly, elevated and depressed. This is in contrast to a ``fossil'' population induced by stellar encounters \citep{ML04}, where $q$ changes are almost instantaneous, or capture of these bodies from a passing star \citep{Jil+15}. 
An excess of classical and large semimajor axis Centaurs,  which are icy bodies with $q$ among the giant planets and $a>100$ au, would be natural in the continuously disturbed population \citep{Gom+15}. 
The influx of comets into the inner solar system would be altered as well \citep{Gom+06,PK16}.

\begin{figure*}
\begin{center}
\resizebox{0.9\hsize}{!}{\includegraphics{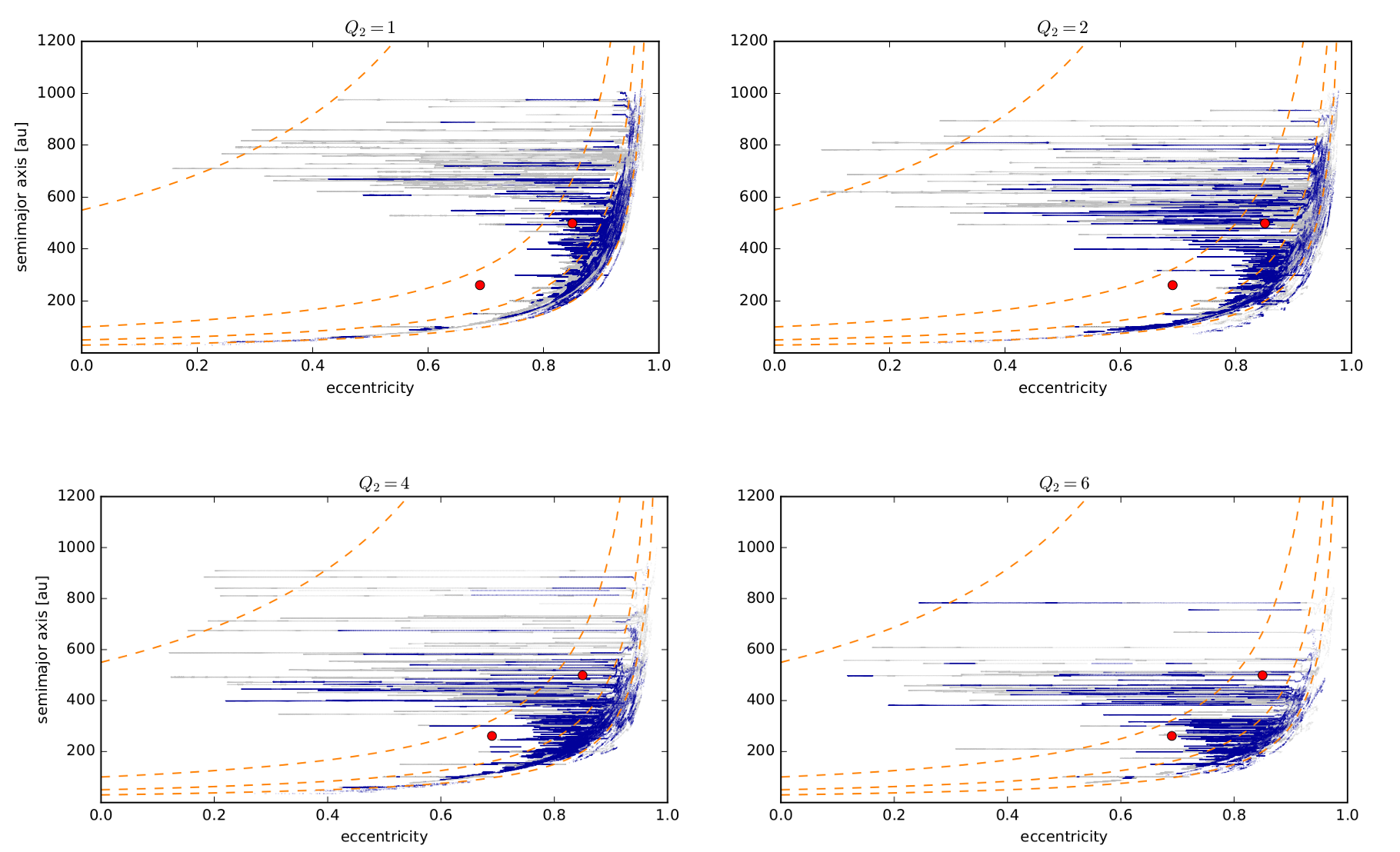}}
\caption{$a-e$ diagram from the last Gyr of the benchmark model simulation incorporating gravitational interaction with Neptune. $Q_{2}=1$ (top left panel), $Q_{2}=2$ (top right panel), $Q_{2}=4$ (bottom left panel), and $Q_{2}=6$ (bottom right panel) were assumed. Two colours (transparency levels) are used as a proxy for two inclination ranges.
The blue (less transparent) evolutionary tracks represent orbits with inclinations spanning $i\in(10,30)$ deg, the interval the inclinations of Sednoids lie in. The grey (more transparent) tracks refer to all the other long-living orbits. Sedna and 2012 VP 113 are indicated as large dots. The dashed lines indicate $q=30$, 50, 100, and 550 au levels (from right/bottom to left/upwards). Objects between the two left-most/upper-most lines ($q$ in range between 100 and 550 au) would hardly be observed if they really exist.}
\label{img:1246}
\end{center}
\end{figure*}

Evolution in $a$--$e$ plane is depicted during the last Gyr of the benchmark model simulation in Fig. \ref{img:1246} for various values of $Q_{2}$. Figures showing $a$--$e$ diagrams are always based on the simulations starting with 800 particles.
With $Q_{2}=0$ and no P9, the particles would be stuck between the two rightmost/bottom-most dashed lines ($q$ in range between 30 and 50 au). Moreover, assuming $Q_{2}=0$ and no P9, inclinations of the particles remain low, typically $i<10$ deg.

From Fig. \ref{img:1246} we can conclude that the orbit of Sedna constrains $Q_{2}$ to $Q_{2}>0.5$. The orbit of 2012 VP$_{113}$, with similar perihelion distance but much lower semimajor axis than Sedna, requires $Q_{2}>4$, if we demand that EFE is solely responsible for the elevation of perihelia. Even with $Q_{2}=4$ -- 6, 2012 VP$_{113}$ still lies at the low-end border of the achievable eccentricities at $a\approx250$ au. Objects with higher eccentricities are easier to detect. The orbit of 2012 VP$_{113}$ is thus only marginally compatible with the EFE origin scenario.

\subsubsection{Beyond the benchmark model}\label{sec:tweak}

What effect can be expected from the next lowest-order correction (octupole) neglected in Eq. (\ref{mond_effect2}) and in the benchmark model? This correction (adds to RHS of Eq. (\ref{mond_effect2})) can be written as\footnote{\cite{BN11} state an opposite sign in the RHS of Eq. (\ref{octupole}), as they have been using $\nabla^{2}\Phi_{N}=-4\pi G \rho$ and ${\bf g}=\nabla\Phi$. We used $\nabla^{2}\Phi_{N}=4\pi G \rho$ and ${\bf g}=-\nabla\Phi$ throughout the paper, hence the difference.} \citep{BN11}
\begin{eqnarray}\label{octupole}
\delta\Phi_{oct} &=& \frac{Q_{3}}{6} r^{i}r^{j}r^{k}\left[e_{i}e_{j}e_{k}-\frac{1}{5}\left(
e_{i}\delta_{jk}+e_{j}\delta_{ik}+e_{k}\delta_{ij}\right)\right]~.
\end{eqnarray}
The octupole strength parameter $Q_{3}$ (typically $Q_{3}<0$) depends on the specific modified Poisson formulation, form of the interpolating function, and strength of the external field in natural units $a_{0}$. \ The parameter
$Q_{3}$ is position dependent in general. The variation of $Q_{3}$ within our volume of interest is up to several percent \citep{BN11}. We treat $Q_{3}$ as position independent, $Q_{3}=Q_{3}(0)$.

\cite{BN11} tested some of the commonly used interpolating functions $\mu,$ for instance the $\mu_{n}(x)=x/(1+x^{n})^{1/n}$ family, in the framework of AQUAL, and found that $\vert Q_{3}\vert < 1.2\times 10^{-40}$ m$^{-1}$ s$^{-2}$ close to the Sun. From now on, when we refer to $Q_{3}$ without explicitly stating its units, the unit is $10^{-40}$ m$^{-1}$ s$^{-2}$. The highest $\vert Q_{3}\vert$ was found for the simple interpolating function $\mu_{1}(x)=x/(1+x)$. 
It is known that $\mu_{1}$, and its inverse $\nu_{1}$, are excluded by the solar system tests \citep{BN11,H+16} because they give a  $Q_{2}$ that is too large.
The higher $n$ functions from the $\mu_{n}$ family, allowed by the solar system data \citep{BN11,H+16}, yield typically $\vert Q_{3}\vert\sim 10^{-2}$ \citep{BN11}. We found that with such a low value of $Q_{3}$, the incorporation of the octupole term, Eq. (\ref{octupole}), has a negligible effect. We note that the most stringent constraints on $Q_{2}$ comes from \cite{H+16}, where QUMOND framework was applied. The values of $Q_{3}$ that we refer to come from \cite{BN11} and were calculated in AQUAL. Thus, when we link the two we assume that their values are similar in both modified gravity frameworks, for a given pair of interpolating functions $\mu$ and $\nu$, and a given external field strength. Approximation of this kind is common; see e.g. \cite{Mil09}.

\begin{figure}
\begin{center}
\resizebox{0.9\hsize}{!}{\includegraphics{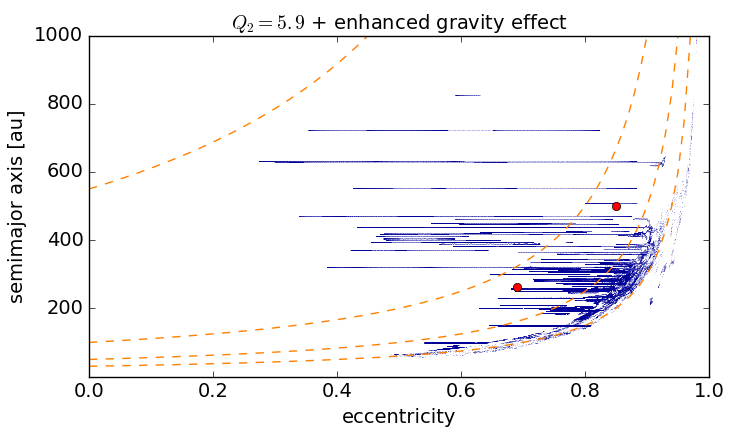}}
\caption{$a-e$ diagram from the last Gyr of the benchmark model simulation incorporating gravitational interaction with Neptune. The enhanced gravity effect is accounted for in the model using the interpolating function $\overline{\nu}_{2}$. $\overline{\nu}_{2}$ yields $Q_{2}=5.9$. Only the evolution of test particles in orbits with inclinations spanning $i\in(10,30)$ deg, which is the interval that the inclinations of Sednoids lie in, is shown. Sedna and 2012 VP 113 are indicated as large dots. The dashed lines indicate $q=30$, 50, 100, and 550 au levels (from right/bottom to left/upwards).}
\label{img:ae_eg}
\end{center}
\end{figure}

Switching back to $\overline{\nu}_{n}$ family and QUMOND, $\overline{\nu}_{2}$ yields $Q_{2}=5.9$ when one assumes $\eta\equiv g_{e}/a_{0}=2.3$, $a_{0}=0.815\times 10^{-10}$ m s$^{-2}$ \citep{H+16}. The quoted value of $a_{0}$ comes from rotation curves fits using $\overline{\nu}_{2}$.
With $\overline{\nu}_{2}$ one gets, besides EFE, as well as a non-negligible enhanced gravity effect. We incorporated the enhanced gravity effect, Eq. (\ref{mond_effect1}), into the model and repeated the simulation. We used $\overline{\nu}_{2}(y)$, $y=\vert {\bf g}^{N}_{\odot}+{\bf g}^{N}_{e}\vert/a_{0}$, where ${\bf g}^{N}_{\odot}$ is Newtonian gravitation from the Sun, and ${\bf g}^{N}_{e}$ was found from ${\bf g}_{e}=\overline{\nu}_{2}(\vert{\bf g}^{N}_{e}\vert/a_{0}){\bf g}^{N}_{e}$ knowing ${\bf g}_{e}$. Direction of fields  ${\bf g}_{e}$ and  ${\bf g}^{N}_{e}$ rotate as prescribed by Eq. (\ref{unit_vector}). An outcome is in Fig. \ref{img:ae_eg}. 2012 VP$_{113}$ still lies at the low-end border of the achievable eccentricities at $a\approx250$ au.
We remind that $\overline{\nu}_{n}$, $n<2$, is disfavoured by the Cassini data \citep{H+16}, and for $\overline{\nu}_{n}$, $n\gtrsim3$, the enhanced gravity effect is completely negligible in the ETNOs region.

\begin{figure}
\begin{center}
\resizebox{0.9\hsize}{!}{\includegraphics{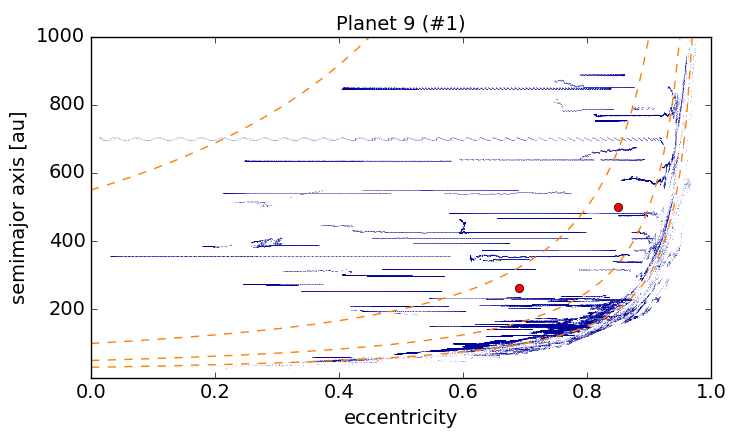}}
\caption{$a-e$ diagram from the last Gyr of the P9 simulation incorporating gravitational interaction with Neptune. Only evolution of test particles in orbits with inclinations spanning $i\in(10,30)$ deg, which is the interval that the inclinations of Sednoids lie in, is shown. Sedna and 2012 VP 113 are indicated as large dots. The dashed lines indicate $q=30$, 50, 100, and 550 au levels (from right/bottom to left/upwards).}
\label{img:aep91}
\end{center}
\end{figure}

We can now directly compare the MD and P9 scenarios. The last Gyr of the Newtonian simulation $(Q_{2}=0)$ with P9 is shown in Fig. \ref{img:aep91} for P9 starting orbit (\#1). Starting in orbit (\#2), P9  yields very similar picture. Apparently, P9 accounts well for the existence of the two. Although orbits close to Sednoids in terms of $a$, $e$, and $i$ were not induced (or not preserved) in the simulation, probably because of the low number of test particles, there were orbits induced with lower semimajor axes and higher perihelion distances than those of Sednoids, which means that the effect of P9 should be sufficient to shape orbits of Sednoids. This is not a new result, P9 was designed to do this.

We also tested how Figs. \ref{img:1246} and \ref{img:ae_eg} change with longer integration time of 4.5 Gyr and found that the differences are minute.
Additionally, we varied also phase $\tau$ and angular speed of the Sun $w$, present in Eq. (\ref{unit_vector}). We used the fact that $w$ was found to lie in bounds $29.8-31.5$ km s$^{-1}$ kpc$^{-1}$ \citep{McM+10}. It reveals that the effect of these variations is of little importance too.

\subsubsection{The case of negative $Q_{2}$}

\begin{figure*}
\begin{center}
\resizebox{0.9\hsize}{!}{\includegraphics{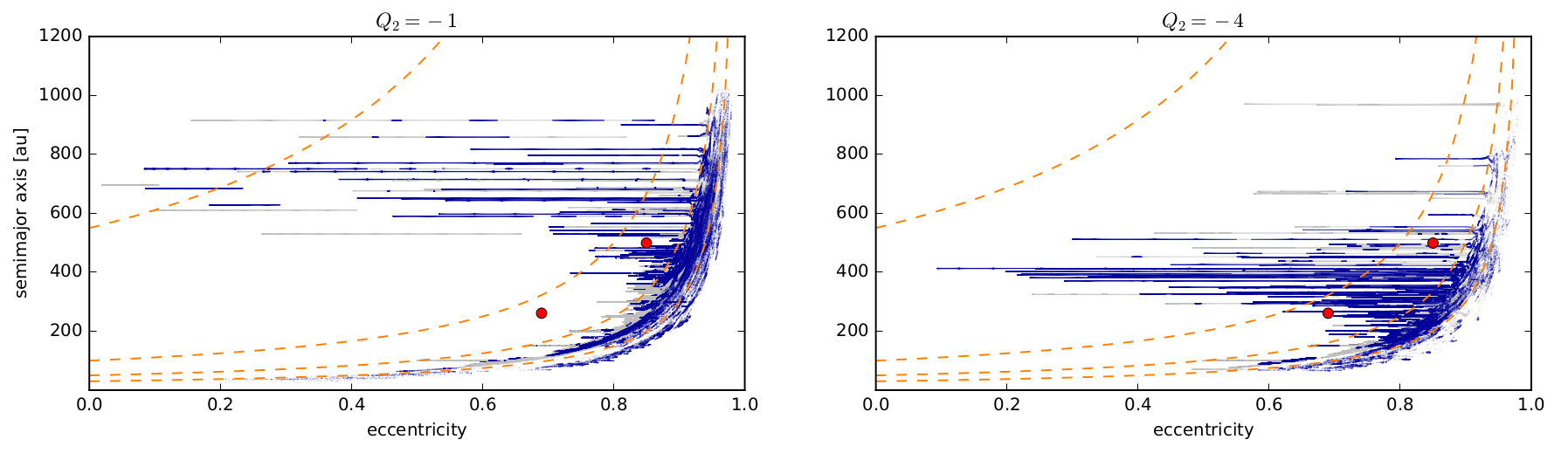}}
\caption{$a-e$ diagram from the last Gyr of the benchmark model simulation incorporating gravitational interaction with Neptune. Negative values of $Q_{2}$, $Q_{2}=-1$ (left panel) and $Q_{2}=-4$ (right panel) were used. Two colours (trasparency levels) are used as a proxy for two inclination ranges. The blue (less transparent) evolutionary tracks represent orbits with inclinations spanning $i\in(10,30)$ deg, which is the interval that the inclinations of Sednoids lie in. The grey (more transparent) tracks refer to all the other long-living orbits. Sedna and 2012 VP 113 are indicated as large dots. The dashed lines indicate $q=30$, 50, 100, and 550 au levels (from right/bottom to left/upwards).}
\label{img:aem14}
\end{center}
\end{figure*}

We show $a-e$ diagrams that are analogous to the previous diagrams, but now with negative values of $Q_{2}$, $Q_{2}=-1,$ and $Q_{2}=-4$, in Fig. \ref{img:aem14}. The orbits of both Sednoids are produced in the MD model as long as $\vert Q_{2}\vert>4$. With $Q_{2}=-4$, 2012 VP$_{113}$ lies at the low-end border of the achievable eccentricities at $a\approx250$ au.

\subsection{Orbital clustering}

Can statistics on ETNOs orbits be understood with the aid of MD without additional planets in the solar system?
At first, we analysed our simulations without explicit visualisation.
We performed the Kolmogorov-Smirnov (KS) test\footnote{To be clear in statistical terms and notation, we give a following example: the probability $p$-value being equal to 0.05 means that the null hypothesis is not rejected at $5\%$ level; in our case the null hypothesis is always the hypothesis in which the two compared datasets have the same distribution.} comparing distributions of $\omega$, $\Omega$, $L$, and $i$ of the observational data with those inferred from the benchmark model simulation (MD model test) at the time instant $t = 4$ Gyr, considering various values of $Q_{2}$.
Additionally, we also did the KS test based on results from $Q_{2}=0$ (no effects of MD in the solar system) simulation with (P9 test) and without (control test) P9. Only test particle orbits fulfilling the criteria -- $i<50$ deg, $q<80$ au at $t=4$ Gyr, were considered to account for the observational preference of such orbits.
The orbits are discerned according to their semimajor axis and divided into two groups: one with $150<a<250$ and the other with $a>250$ au. The results are presented in Table \ref{tab:KS_test1} for simulations starting with 400 particles.

\begin{table}[]
\centering
\caption{Results of the KS test comparing distributions of $\omega$, $\Omega$, $L$, and $i$ of the observational data and those inferred from the benchmark model simulation at $t = 4$ Gyr. The KS test was also carried out for the P9 simulation. Each simulation starts with 400 particles; $p(\xi~|~\zeta)$ is the probability p-value, where $\xi$ are the observational data in a quoted $a$-range, and $\zeta$ are the simulated data in a quoted $a$-range, e.g. $p(150 - 250~|~150 - 250)$ is the p-value comparing the observational and simulated data in range $150 < a < 250$ au. The $p$-values larger than 0.5 are highlighted in boldface. There was only one particle with $a>250$ au left at the end of the P9 simulation, therefore $p(>250~|>250)$ is not reported in this case;
$[Q_{2}]$ $=$ $10^{-27}$ s$^{-2}$.}
\label{tab:KS_test1}
\begin{tabular}{|c|cccc|}
\hline
$Q_{2}=0$ & $\omega$ & $\Omega$ & $L$ & $i$  \\
\hline
$p(150-250~|~150-250)$   &  0.17  &  0.31  &  \textbf{0.66}  &  0.00  \\
$p(>250~|>250)$         &  0.00  &  0.03  &  0.04  &  0.00  \\
\hline
\hline
P9 (\#1) & $\omega$ & $\Omega$ & $L$ & $i$  \\
\hline
$p(150 - 250~|~150 - 250)$  &  \textbf{0.95}  &  0.25  &  0.07  &  \textbf{0.98} \\
\hline
\hline
P9 (\#2) & $\omega$ & $\Omega$ & $L$ & $i$  \\
\hline
$p(150 - 250~|~150 - 250)$  &  0.48  &  0.48  &  \textbf{0.83}  &  \textbf{1.00} \\
\hline
\hline
$Q_{2}=1$ & $\omega$ & $\Omega$ & $L$ & $i$  \\
\hline
$p(150-250~|~150-250)$   &  0.19  &  0.03  &  \textbf{0.71}  &  0.00  \\
$p(>250~|>250)$         &  0.02  &  0.00  &  0.01  &  0.00  \\
\hline
\hline
$Q_{2}=2$ & $\omega$ & $\Omega$ & $L$ & $i$  \\
\hline
$p(150-250~|~150-250)$   &  0.25  &  0.01  &  \textbf{0.95}  &  0.00  \\
$p(>250~|>250)$         &  0.01  &  0.01  &  0.01  &  0.02  \\
\hline
\hline
$Q_{2}=4$ & $\omega$ & $\Omega$ & $L$ & $i$  \\
\hline
$p(150-250~|~150-250)$   &  0.35  &  0.00  &  \textbf{0.66}  &  0.00  \\
$p(>250~|>250)$         &  0.03  &  0.00  &  0.04  &  0.49  \\
\hline
\hline
$Q_{2}=6$ & $\omega$ & $\Omega$ & $L$ & $i$  \\
\hline
$p(150-250~|~150-250)$   &  0.26  &  0.01  &  \textbf{0.80}  &  0.07  \\
$p(>250~|>250)$         &  0.00  &  0.00  &  0.01  &  \textbf{0.89}  \\
\hline
\end{tabular}
\end{table}

As there was only one test particle with $a>250$ au left in the P9 simulation at $t=4$ Gyr (for both considered initial orbits of P9), $p$-values, $p(>250~|>250)$, are not reported in the P9 case.
The P9 test shows that an incorporation of P9 in the Newtonian model increases its likelihood for $a<250$ au, though the compared samples are small ($\sim$10).
A possible caveat of the P9 scenario was revealed by \cite{Law+17}. They performed a similar simulation starting with many more test particles $(10^{5})$, interacting gravitationally with five planets (known giant planets + P9) in a full N-body manner for 4 Gyr, leading to no angular clustering at the end of the simulation.

In the semimajor axis range $150<a<250$ au, the MD model test yields $p$-values, $p(150-250~|~150-250)$, similar to the control test, except for the distribution of $\Omega$ when even lower $p$-values are reported. This means some clustering in $\Omega$ in the simulated data in MD but at different positions than the observed data.

At the $a>250$ au level, the observed distribution of $i$ might be compatible with EFE if $Q_{2}\gtrsim 4$. However, the compatibility between the observational data and the MD model remains low in the case of $\omega$ and $L$, though $p$-values, $p(>250~|>250)$, are slightly higher than in the control test if, e.g. $Q_{2}=4$. The distribution of $\Omega$ seems to be the most problematic distribution, with the simulated data clustered but having significant offset from the observed clustering position.
 
We tested how the results in Table \ref{tab:KS_test1} change when we restrict the orbits of particles to those with $q>35$ au. An example is given in Table \ref{tab:KS_qr35}. The difference is small. There is prevalence of $q>35$ au orbits at $t=4$ Gyr. Doubling the number of test particles at $t=0$ does not change the results qualitatively. An example can be seen in Table \ref{tab:KS_800}. Even after doubling the initial number of particles, there was only one particle with $a>250$ au left at the end of the P9 simulation for both considered starting orbits of P9.

\begin{table}[]
\centering
\caption{Results of the KS test taking into account only the orbits, of ETNOs and test particles, with $q>35$ au.}
\label{tab:KS_qr35}
\begin{tabular}{|c|cccc|}
\hline
P9 (\#1) & $\omega$ & $\Omega$ & $L$ & $i$  \\
\hline
$p(150 - 250~|~150 - 250)$  &  \textbf{0.71}  &  \textbf{0.71}  &  0.28  &  \textbf{0.53} \\
\hline
\hline
$Q_{2}=1$ & $\omega$ & $\Omega$ & $L$ & $i$  \\
\hline
$p(150-250~|~150-250)$   &  0.11  &  0.18  &  \textbf{0.66}  &  0.00  \\
$p(>250~|>250)$         &  0.02  &  0.00  &  0.01  &  0.00  \\
\hline
\hline
$Q_{2}=6$ & $\omega$ & $\Omega$ & $L$ & $i$  \\
\hline
$p(150-250~|~150-250)$   &  0.33  &  0.02  &  \textbf{0.99}  &  0.22  \\
$p(>250~|>250)$         &  0.01  &  0.00  &  0.01  &  \textbf{0.64}  \\
\hline
\end{tabular}
\end{table}

\begin{table}[]
\centering
\caption{Results of the KS test after doubling the initial number of test particles. We only considered the orbits of ETNOs and test particles with $q>35$ au. There was only one particle with $a>250$ au left at the end of the P9 simulation, therefore $p(>250~|>250)$ is not reported in the P9 case.}
\label{tab:KS_800}
\begin{tabular}{|c|cccc|}
\hline
P9 (\#1) & $\omega$ & $\Omega$ & $L$ & $i$  \\
\hline
$p(150 - 250~|~150 - 250)$  &  \textbf{0.98}  &  \textbf{0.57}  &  \textbf{0.98}  &  \textbf{0.57} \\
\hline
\hline
$Q_{2}=1$ & $\omega$ & $\Omega$ & $L$ & $i$  \\
\hline
$p(150-250~|~150-250)$   &  0.09  &  0.27  &  \textbf{0.85}  &  0.00  \\
$p(>250~|>250)$         &  0.01  &  0.00  &  0.01  &  0.00  \\
\hline
\hline
$Q_{2}=6$ & $\omega$ & $\Omega$ & $L$ & $i$  \\
\hline
$p(150-250~|~150-250)$   &  0.19  &  0.09  &  \textbf{0.86}  &  0.01  \\
$p(>250~|>250)$         &  0.00  &  0.00  &  0.03  &  \textbf{0.51}  \\
\hline
\end{tabular}
\end{table}

\begin{figure}
\begin{center}
\resizebox{0.9\hsize}{!}{\includegraphics{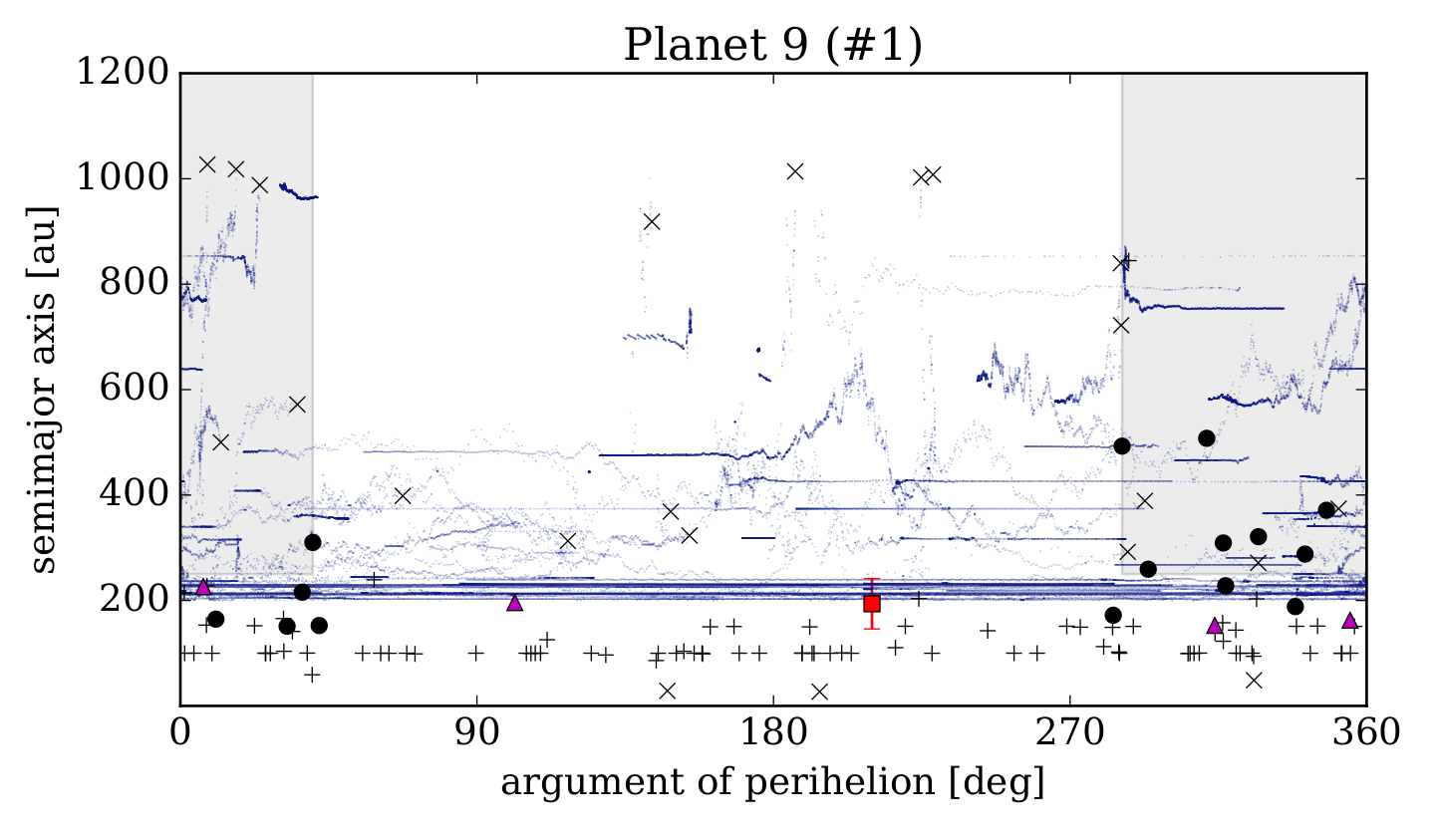}}
\caption{Evolution of test particle orbits, taken from the P9 simulation, in $\omega-a$ plane. Only particles surviving at least 3 Gyr are taken into account. The evolutionary tracks of the particles are shown in the last Gyr of the simulation every time their orbits met the criteria: $i<50$ deg, $q<80$ au. The tracks below $a=200$ au are not shown as these are just a forest of full-width horizontal lines.
The X symbols indicate the positions of potentially unstable particles, which are discarded at some time between $t=3$ and 4 Gyr, at the time of their destruction. The plus signs indicate positions of long-living particles at $t=4$ Gyr.
The known ETNOs with $q>35$ au (circles) and $q<35$ au (triangles) are plotted. TNO 2003 SS$_{422}$ is denoted as a square and its large uncertainty in $a$ is indicated by the error bar. Shaded regions indicate the observed clustering regions for $a>250$ au ETNOs.}
\label{img:p91_om}
\end{center}
\end{figure}

\begin{figure*}
\begin{center}
\resizebox{0.9\hsize}{!}{\includegraphics{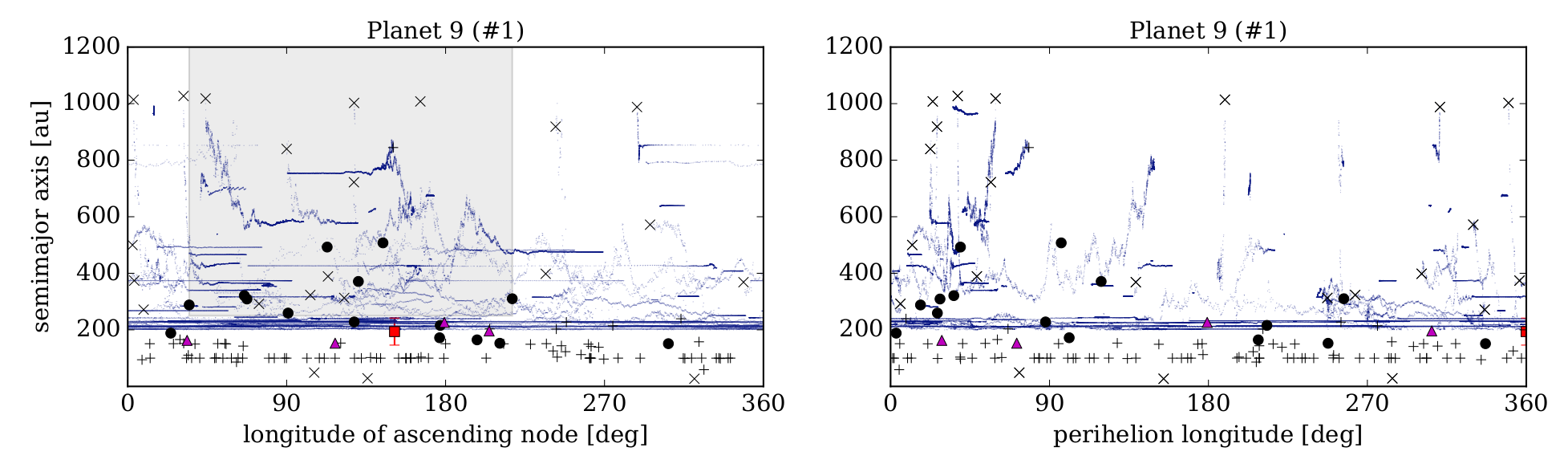}}
\caption{Evolution of test particle orbits, taken from the P9 simulation, in $\Omega-a$ (left panel) and $L-a$ plane (right panel).}
\label{img:p91_Omg_L}
\end{center}
\end{figure*}

We now investigate the simulated data visually. Figures \ref{img:p91_om} - \ref{img:p91_Omg_L} show evolution of test particle orbits in $\omega-a$, $\Omega-a$, and $L-a$ plane in the last Gyr of the Newtonian simulation with P9. We initialised P9 in the orbit (\#1). Initialising P9 in the orbit (\#2) would have yielded a similar picture. Evolutionary tracks of the particles are shown every time their orbits met the criteria, i.e. $i<50$ deg, $q<80$ au. 
The evolutionary tracks continue down to $a=100$ au, but we do not show them below $a=200$ au as they are just a forest of full-width horizontal lines.
Clustering in $\omega$ emerges only for $a\gtrsim500$ au, as could be seen already in \cite{BB16}. Potentially unstable particles, discarded at some time between $t=3$ and 4 Gyr, have $\omega$ preferentially close to 0 (360) or 180 deg at the time of their destruction, which is another feature already noticed by \cite{BB16b}. The relative number of particles with $\omega\approx0$ to the number of particles with $\omega\approx180$ can be influenced by altering an initial orbit of Planet 9 \citep{BB16b}. The best agreement between observations and the P9 model can be seen in terms of $L$. However, the clustering in orbital angles is always traced only by the potentially unstable particles.

\begin{figure}
\begin{center}
\resizebox{0.9\hsize}{!}{\includegraphics{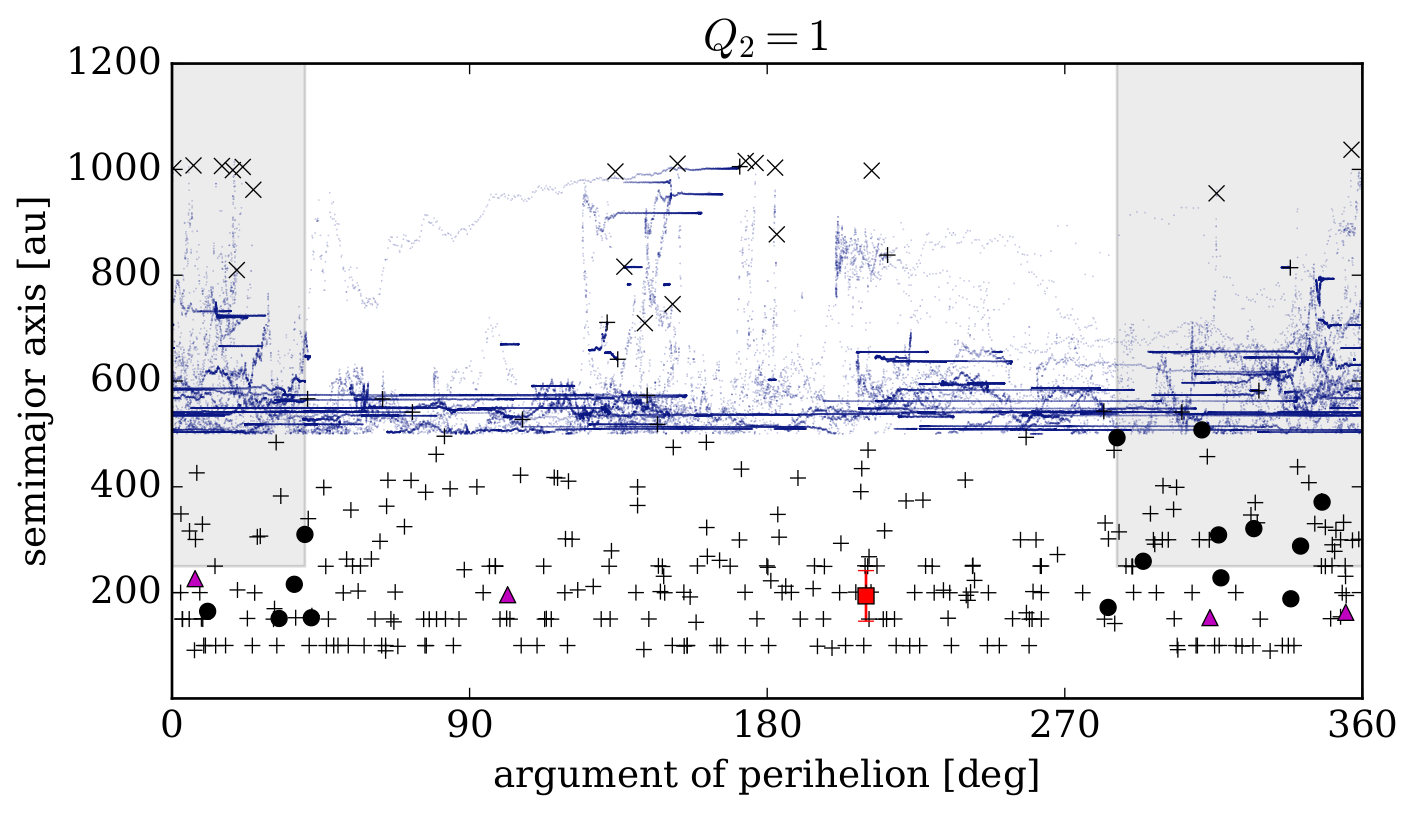}}
\caption{Evolution of test particle orbits, taken from the benchmark model simulation, in $\omega-a$ plane. The strength of EFE was assumed to be $Q_{2}=1$. Only particles surviving at least 3 Gyr are taken into account. The evolutionary tracks of the particles are shown in the last Gyr of the simulation every time their orbits met the criteria, i.e. $i<50$ deg, $q<80$ au. The tracks below $a=500$ au are not shown as these are just full-width horizontal lines.}
\label{img:Q1_om}
\end{center}
\end{figure}

\begin{figure*}
\begin{center}
\resizebox{0.9\hsize}{!}{\includegraphics{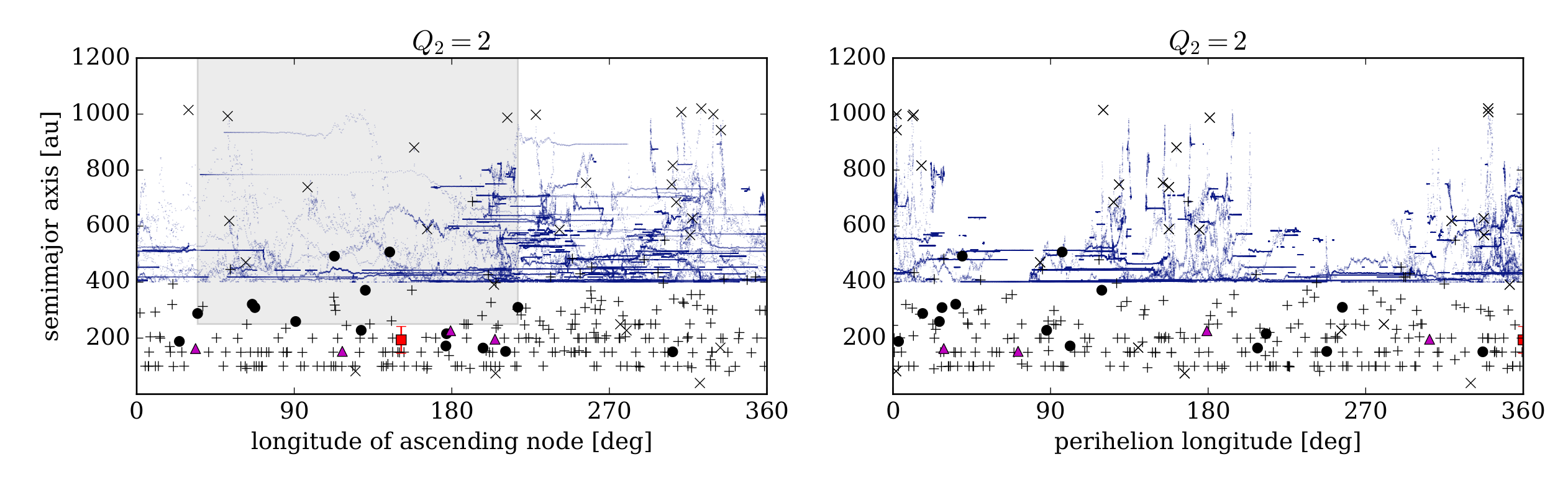}}
\caption{Evolution of test particle orbits, taken from the benchmark model simulation, in $\Omega-a$ (left panel) and $L-a$ (right panel) plane. The strength of EFE was assumed to be $Q_{2}=2$. Only particles surviving at least 3 Gyr are taken into account. Evolutionary tracks of the particles are shown in the last Gyr of the simulation every time their orbits met the criteria, i.e. $i<50$ deg, $q<80$ au. The tracks below $a=400$ au are not shown as these are just full-width horizontal lines. Increasing $Q_{2}$, the clustering in $L$ sets in at lower semimajor axes, e.g. with $Q_{2}=6$; similar clustering as in the figure develops for $a>300$ au.}
\label{img:Q2_Omg_L}
\end{center}
\end{figure*}

We turn our attention to MD.
Figure \ref{img:Q1_om} shows the evolution of test particle orbits in $\omega-a$ plane in the last Gyr of the benchmark model simulation assuming $Q_{2}=1$. Again, only particles surviving at least 3 Gyr, orbiting in the observationally preferred orbits, are taken into account. Similarly with the P9 picture, clustering in $\omega$ develops only for large $a$ orbits, in the case of $Q_{2}=1$, $a>600$ au,  and it is only traced by the potentially unstable orbits. The value of $\omega$ near 0 (360) as well as near 180 deg is preferred by the test particles. With $Q_{2}\gtrsim 2$, the clustering in $\omega$ fades away.

The test particle orbits strongly cluster around $\Omega\approx270$ deg. This is illustrated in the left panel of Fig. \ref{img:Q2_Omg_L}, where we show results from $Q_{2}=2$ simulation. Strong confinement in $\Omega$ develops for stable orbits in a wide range of $Q_{2}$ values, $Q_{2}\in(0.5,6)$, and also for potentially unstable orbits if $Q_{2}\gtrsim2$.

The distribution of $L$ shows a coherent picture for $Q_{2}$ in the range $Q_{2}\in(1,6)$. Clustering in $L$ is noticeable around $L=0$ (360) and $L=180$ deg and is traced again mainly by potentially unstable orbits. In the right panel of Fig. \ref{img:Q2_Omg_L}, we show the evolution of test particle orbits in $L-a$ plane, as taken from $Q_{2}=2$ simulation. As $Q_{2}$ increases, clustering develops at lower and lower semimajor axes. For example, for $Q_{2}=1$, the clustering develops only beyond $a=500$ au, while the border is at $a=300$ au for $Q_{2}=6$.

\begin{figure*}
\begin{center}
\resizebox{0.85\hsize}{!}{\includegraphics{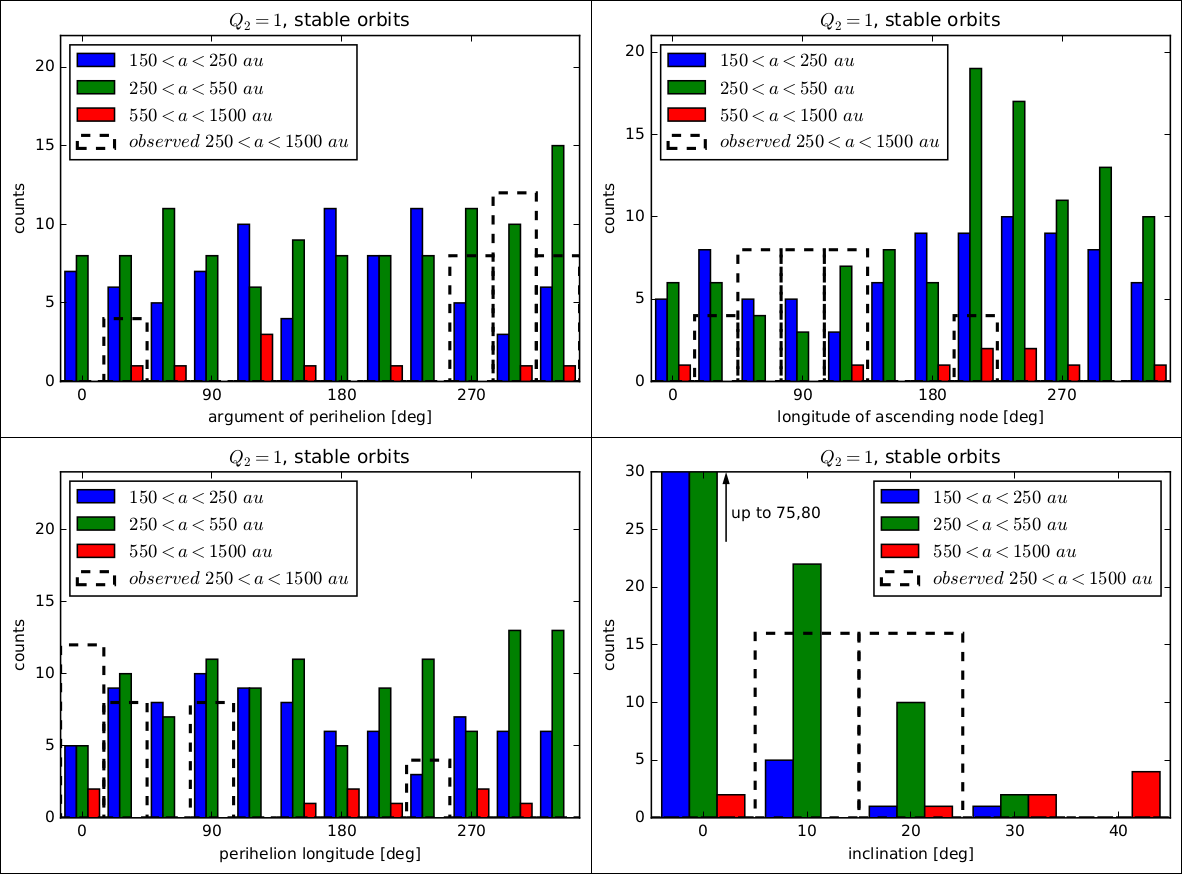}}
\caption{Histogram distributions of $\omega$ (top left panel), $\Omega$ (top right panel), $L$ (bottom left panel), and $i$ (bottom right panel) of test particle orbits at the end of the benchmark model simulation, i.e. at $t=4$ Gyr. The strength of EFE was assumed to be $Q_{2}=1$. Orbits are discerned according to their semimajor axis and divided into three groups: $150<a<250$ au (blue histograms), $250<a<550$ au (green histograms), and $550<a<1500$ au (red histograms). The observed distributions for ETNOs with $250<a<1500$ au, multiplied by factor 4, are shown as well (dashed histograms). Histograms are aligned to the left.}
\label{img:1_hist}
\end{center}
\end{figure*}

\begin{figure*}
\begin{center}
\resizebox{0.85\hsize}{!}{\includegraphics{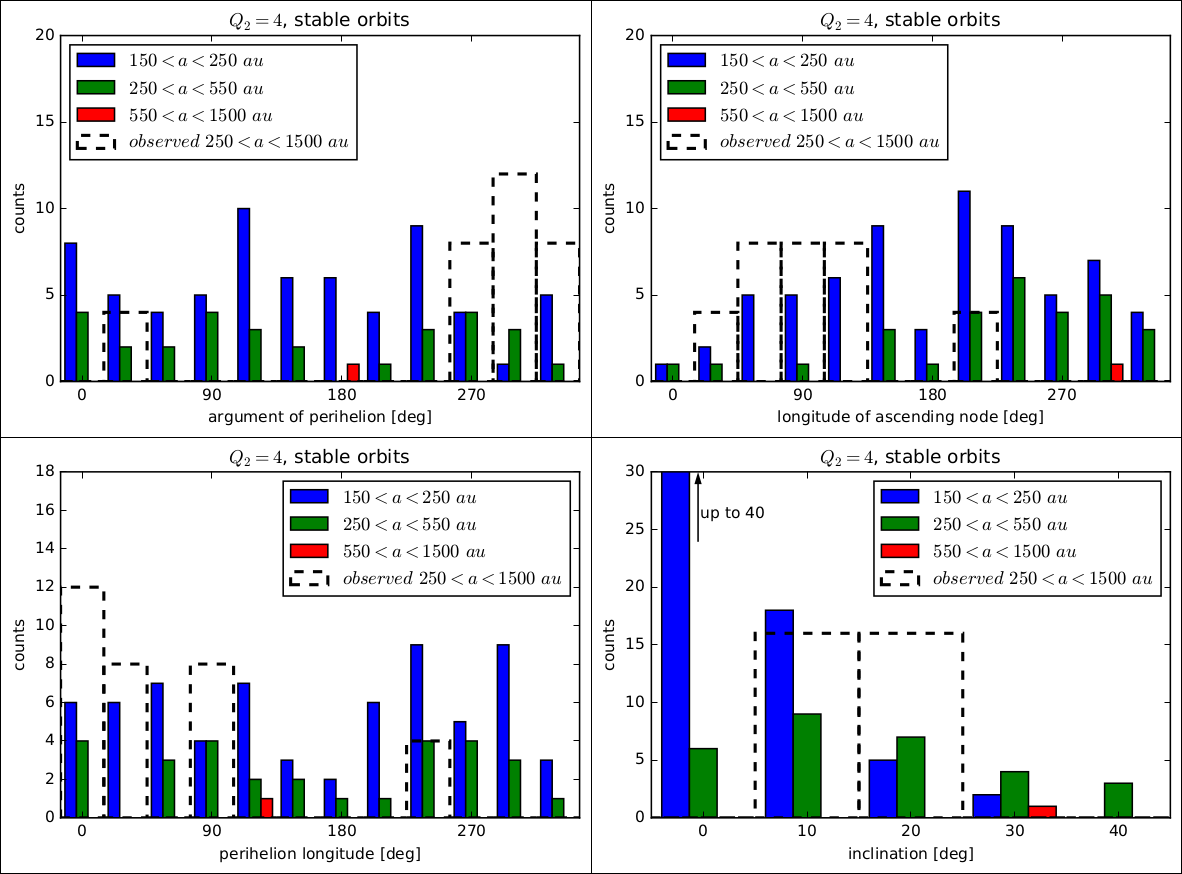}}
\caption{Histogram distributions of $\omega$ (top left panel), $\Omega$ (top right panel), $L$ (bottom left panel), and $i$ bottom right panel) of test particle orbits at the end of the benchmark model simulation, i.e. at $t=4$ Gyr. The strength of EFE was assumed to be $Q_{2}=4$.}
\label{img:4_hist}
\end{center}
\end{figure*}

We depict histogram distributions of $\omega$, $\Omega$, $L$, and $i$ of test particle orbits at the end of the simulation, i.e. at $t=4$ Gyr, in Figs. \ref{img:1_hist} ($Q_{2}=1$) and \ref{img:4_hist} ($Q_{2}=4$). Only particles with $a>150$ au and $q>35$ au are considered. Stable particles show no apparent clustering in terms of $\omega$. Strong clustering emerges in terms of $\Omega$ across wide range of $Q_{2}$ values, $Q_{2}\in(0.5,6)$, as majority of the orbits prefer $\Omega\in(180,360)$ deg at $t=4$ Gyr. This is the trend already identified in the simple numerical exploration of Sec. \ref{sec:model0}: with positive $Q_{2}$, we get a simulated clustering position shifted by 180 deg with respect to the observed clustering position. 

Stable orbits also slightly prefer $L$ around 90 and 270 deg if $Q_{2}\approx4$. This distinction between stable and unstable orbits could be linked to the distinction between observed $q>35$ au and $q<35$ au orbits, looming for $a>150$ objects \citep{ST16}.
By increasing the value of $Q_{2}$, one gets rid of low-inclination ETNOs gradually. Objects with higher inclinations are more resilient. With $Q_{2}= 4 - 6$, majority of the particles in $a>250$ au orbits have $i\in(10,30)$ deg at $t = 4$ Gyr, as observed. But the simulated inclination distribution still does not account for the lack of the observed ETNOs with inclinations lower than 10 deg, bearing in mind their observational preference.

\begin{figure}
\begin{center}
\resizebox{0.85\hsize}{!}{\includegraphics{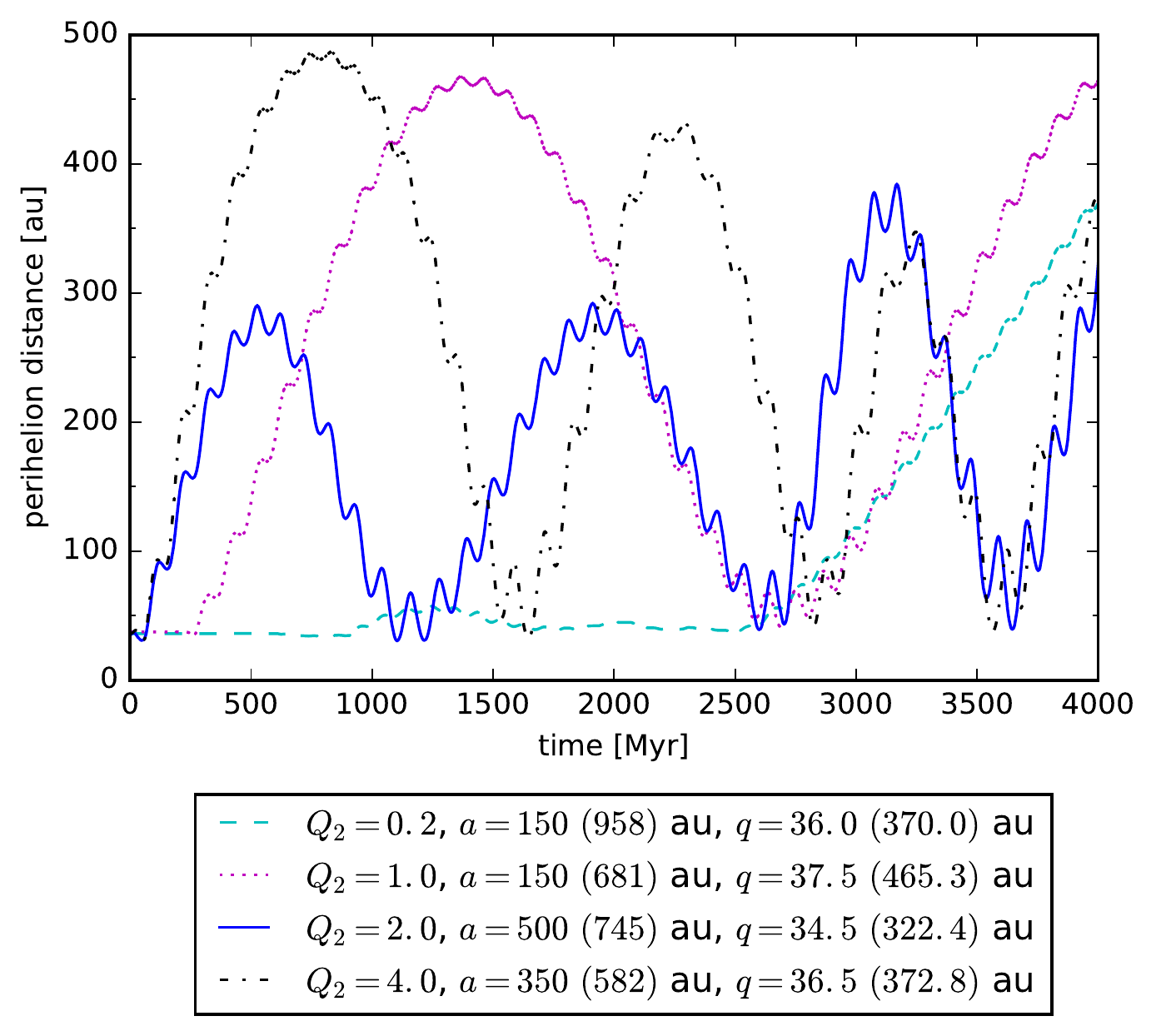}}
\caption{Perihelion distance as a function of time for 4 selected orbits from the benchmark model simulation characterised by high-$q$ amplitude. Assumed values of $Q_{2}$ and initial (final) values of $a$ and $q$ are indicated.}
\label{img:qs}
\end{center}
\end{figure}

Perihelion distance as a function of time for four test particle orbits from the benchmark model simulation, which are characterised by a high amplitude of migration in $q$, is depicted in Fig. \ref{img:qs}. This example is relevant also for a more complicated dynamical model, where both EFE and a distant massive planet that is placed on its orbit with the aid of EFE act on ETNOs. For example, EFE with $Q_{2}=0.2$ could lift the perihelion of P9 from the edge of the planetary region\footnote{We did not examine how migration through the giant planets' region, whence P9 would probably originate from, proceeds. The drawback can be the slowness of the EFE induced migration, as P9 in an eccentric orbit crossing orbits of the giant planets, is endangered by their energetic kicks.} to few hundred au, while with P9 in the orbit (\#1) and EFE of strength $Q_{2}=0.2$, we get similar picture of the ETNOs region as in Figs. \ref{img:p91_om} - \ref{img:p91_Omg_L} with pure P9 model. We do not develop these ideas further in this paper.

\subsubsection{Beyond the benchmark model}\label{sec:tweak2}

We tested the robustness of the $a-e$ diagrams inferred from the benchmark model simulation in Sec. \ref{sec:tweak}.
We have carried out similar robustness tests again, paying attention to the orbital clustering. These tests include the incorporation of the octupole term, Eq. (\ref{octupole}), in the governing potential, assuming $Q_{3}\sim10^{-2}$; the inclusion of the enhanced gravity effect in the case\footnote{We compared this with the benchmark model $Q_{2}=6$ simulation.} $Q_{2}=5.9$; the variation of $w$ and $\tau$ in Eq. (\ref{unit_vector}); and the enhancement of the simulation time to 4.5 Gyr. None of these changes of the model matters; the results remain qualitatively the same.

\subsubsection{The case of negative $Q_{2}$}\label{sec:negativeQ2}

We carried out the KS test that compares distributions of $\omega$, $\Omega$, $L$, and $i$, of simulated and observed ETNOs. The $p$-values of the test are listed in Table \ref{tab:KS_test_neg}.
It is analogous to Table \ref{tab:KS_test1}, but now, MD models with $Q_{2}<0$ are investigated. We restricted ourselves to $\vert Q_{2}\vert<6$ (within $3\sigma$ range given by the Casssini data). In the case of $\omega$, $L$, and $i$ distributions, the reported $p$-values are similar to their $Q_{2}>0$ counterparts as long as models with the same $\vert Q_{2}\vert$ are compared. In the case of $\Omega$, the $p$-values are much higher.

\begin{figure*}
\begin{center}
\resizebox{\hsize}{!}{\includegraphics{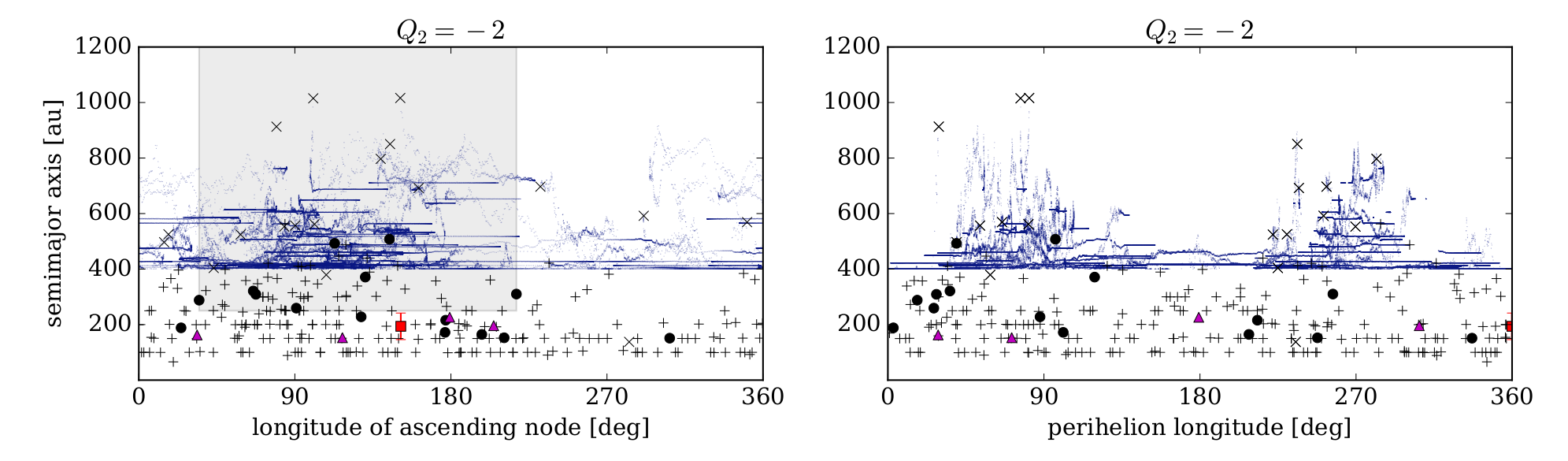}}
\caption{Evolution of test particle orbits, taken from the benchmark model simulation in $\Omega-a$ (left panel) and $L-a$ (right panel) plane. A negative $Q_{2}$, $Q_{2}=-2$ was assumed. Only particles surviving at least 3 Gyr are taken into account. Evolutionary tracks of the particles are shown in the last Gyr of the simulation every time their orbits met the criteria, i.e. $i<50$ deg, $q<80$ au. The tracks below $a=400$ au are not shown as these are just full-width horizontal lines. Decreasing $Q_{2}$, $Q_{2}<0$, the clustering in $L$ sets in at lower semimajor axes, e.g. with $Q_{2}=-6$ similar clustering as in the figure develops for $a>250$ au.}
\label{img:Qm2_Omg_L}
\end{center}
\end{figure*}

Evolution in $\Omega-a$ and $L-a$ plane, inferred from the last Gyr of $Q_{2}=-2$ simulation, is visualised in Fig. \ref{img:Qm2_Omg_L}. Varying $Q_{2}$ in the range $Q_{2}\in(-6,-1)$ yields a qualitatively similar picture as in Fig. \ref{img:Qm2_Omg_L}. Potentially unstable orbits cluster around $L\approx90$ and 270 deg. The clustering in $L$ sets in for $Q_{2}=-1$ beyond $a=450$ au, while for $Q_{2}=-6$ already beyond $a=300$ au. We found no apparent clustering in $\omega$ in our simulations with $Q_{2}<0$.

\begin{figure*}
\begin{center}
\resizebox{0.85\hsize}{!}{\includegraphics{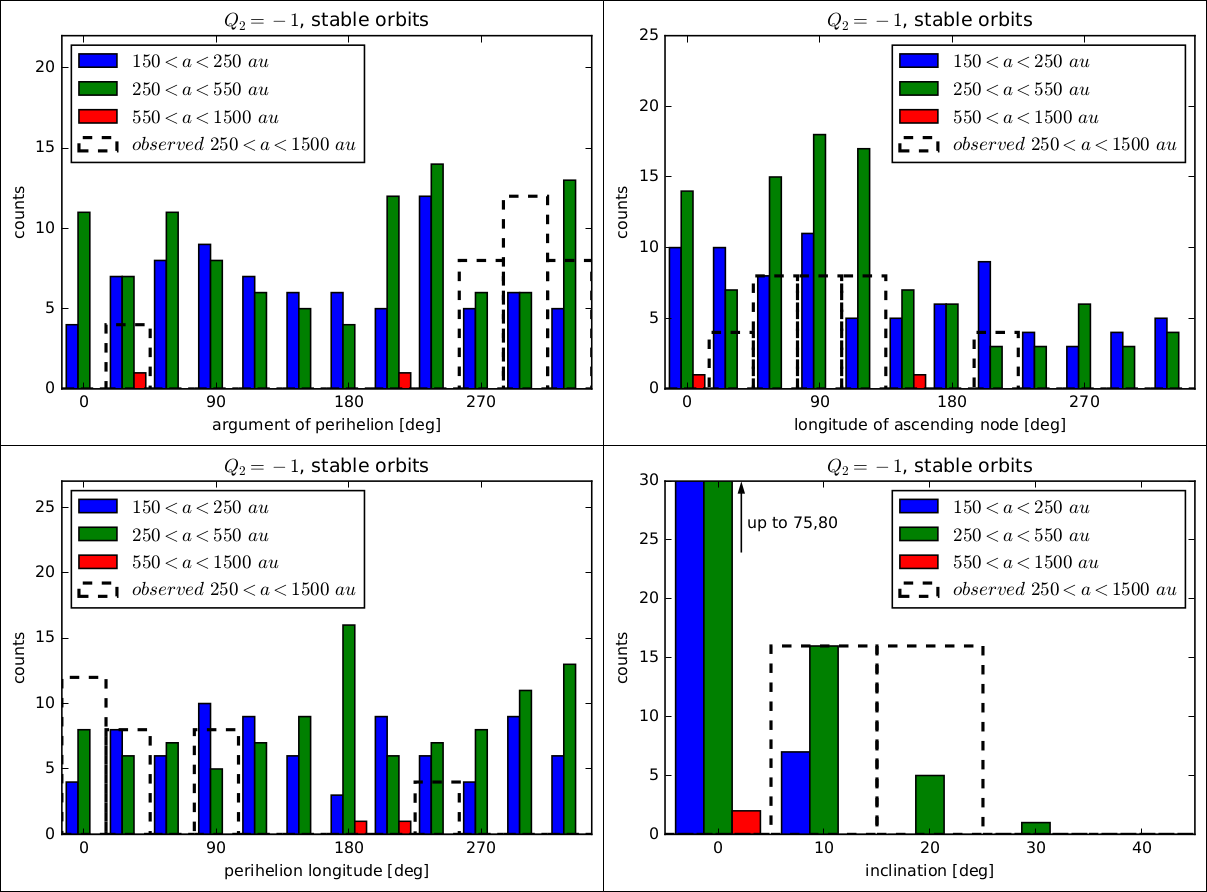}}
\caption{Histogram distributions of $\omega$ (top left panel), $\Omega$ (top right panel), $L$ (bottom left panel), and $i$ (bottom right panel) of test particle orbits at the end of the benchmark model simulation, i.e. at $t=4$ Gyr. A negative $Q_{2}$, $Q_{2}=-1$ was assumed. Orbits are discerned according to their semimajor axis and divided into three groups: $150<a<250$ au (blue histograms), $250<a<550$ au (green histograms), and $550<a<1500$ au (red histograms). The observed distributions for ETNOs with $250<a<1500$ au, multiplied by factor 4, are shown as well (dashed histograms). Histograms are aligned to the left.}
\label{img:m1_hist}
\end{center}
\end{figure*}

Histogram distributions of $\omega$, $\Omega$, $L$, and $i$ inferred at $t=4$ Gyr from the $Q_{2}=-1$  simulation are shown in Fig. \ref{img:m1_hist}.
The clustering in $\Omega$ is now aligned with the observed clustering position and traced by both stable and potentially unstable orbits. With $\vert Q_{2}\vert = 4 - 6$, $Q_{2}<0$, the majority of the particles in $a>250$ au orbits have $i\in(10,30)$ deg at $t = 4$ Gyr, as observed. However, this range of $Q_{2}$ values is highly improbable (probability $<1\%$) in light of the Cassini data.

\begin{table}[]
\centering
\caption{Results of the KS test comparing distributions of $\omega$, $\Omega$, $L$, and $i$ of the observational data and those inferred from the benchmark model simulation at $t = 4$ Gyr. $Q_{2}$ is assumed to be negative, fulfilling $\vert Q_{2}\vert<6$.
$[Q_{2}]$ $=$ $10^{-27}$ s$^{-2}$.}
\label{tab:KS_test_neg}
\begin{tabular}{|c|cccc|}
\hline
$Q_{2}=0$ & $\omega$ & $\Omega$ & $L$ & $i$  \\
\hline
$p(150-250~|~150-250)$   &  0.17  &  0.31  &  \textbf{0.66}  &  0.00  \\
$p(>250~|~>250)$         &  0.00  &  0.03  &  0.04  &  0.00  \\
\hline
\hline
P9 (\#1) & $\omega$ & $\Omega$ & $L$ & $i$  \\
\hline
$p(150 - 250~|~150 - 250)$  &  \textbf{0.95}  &  0.25  &  0.07  &  \textbf{0.98} \\
\hline
\hline
P9 (\#2) & $\omega$ & $\Omega$ & $L$ & $i$  \\
\hline
$p(150 - 250~|~150 - 250)$  &  0.48  &  0.48  &  \textbf{0.83}  &  \textbf{1.00} \\
\hline
\hline
$Q_{2}=-1$ & $\omega$ & $\Omega$ & $L$ & $i$  \\
\hline
$p(150-250~|~150-250)$   &  0.15  &  0.15  &  \textbf{0.81}  &  0.00  \\
$p(>250~|>250)$         &  0.00  &  \textbf{0.85}  &  0.01  &  0.00  \\
\hline
\hline
$Q_{2}=-2$ & $\omega$ & $\Omega$ & $L$ & $i$  \\
\hline
$p(150-250~|~150-250)$   &  0.19  &  0.27  &  \textbf{0.98}  &  0.00  \\
$p(>250~|>250)$         &  0.02  &  \textbf{0.95}  &  0.07  &  0.00  \\
\hline
\hline
$Q_{2}=-4$ & $\omega$ & $\Omega$ & $L$ & $i$  \\
\hline
$p(150-250~|~150-250)$   &  0.30  &  0.48  &  \textbf{0.65}  &  0.01  \\
$p(>250~|>250)$         &  0.00  &  \textbf{0.90}  &  0.04  &  0.19  \\
\hline
\hline
$Q_{2}=-6$ & $\omega$ & $\Omega$ & $L$ & $i$  \\
\hline
$p(150-250~|~150-250)$   &  0.19  &  \textbf{0.53}  &  \textbf{0.66}  &  0.01  \\
$p(>250~|>250)$         &  0.01  &  \textbf{0.68}  &  0.01  &  0.40  \\
\hline
\end{tabular}
\end{table}

\section{Summary and discussion}\label{sec:discuss}

Solar system bodies in orbits with perihelia beyond Neptune and semimajor axes $150<a<1500$ au, dubbed extreme trans-Neptunian objects (ETNOs), have their orbits oriented in space in highly non-isotropic fashion. A general trend in numerical simulations is that any primordial clustering of ETNOs orbits is quickly washed away by the perturbations of giant planets. If the distributions of ETNOs angular orbital elements $\omega$, $\Omega$, and $i$ are not mainly products of an observational bias, then these statistics suggest that there should exist some external perturbation that acts continuously on ETNOs and forces the observed orbital clustering. A natural first guess is that the external perturbation comes from an unseen planet. It turns out that such a planet -- Planet 9 (P9)-- has to be massive $(M\sim 10~M_{\oplus})$; it has to orbit the Sun in a distant, eccentric, and moderately inclined orbit ($a\sim 400 - 700$ au, $q\sim 250$ au, $i\sim 30$ deg) \citep{BB16,BB16b}. Moreover, it has to be far from perihelion right now \citep{Fie+16}. In our contribution, we investigated the hypothesis that the external perturbation does not come from a massive planet in Newtonian dynamics, but from a subtle effect of more general theory of Milgromian dynamics (MD). The dominant effect of MD (formulated as modified gravity) in the ETNOs region is the external field effect (EFE); the external gravitational acceleration from the Galaxy\footnote{Do not confuse this with the very different tidal effect, arising from differential gravity. Existence of EFE means that the strong equivalence principle is broken.} does not decouple from the internal dynamics,  it enters the equation of motion. We modelled EFE in the solar system as a quadrupole correction to the Newtonian potential along the direction of the external Galactic field; see Eq. (\ref{mond_effect2}). We accounted for a variation of the direction of the external field with time. The strength of EFE at a given position was parametrised with the parameter $Q_{2}$, and constrained by the Earth-Cassini range measurements to be $Q_{2}=3\pm3\times 10^{-27}$ s$^{-2}$ (nominal $\pm 1\sigma$ value) \citep{H+14}. 
The action of the giant planets was incorporated into the model. 
Starting with isotropically oriented orbits of primordial ETNOs, we numerically investigated the evolution of these orbits during the subsequent 4 Gyr, considering various values of the parameter $Q_{2}$.

\noindent
Our results can be summarised as follows: 
\begin{itemize}
\item The orbit of Sedna can be explained by EFE torquing if $\vert Q_{2}\vert\gtrsim1\times 10^{-27}$ s$^{-2}$.

\item The orbit of the other Sednoid, 2012 VP$_{113}$, seems only marginally compatible with the EFE scenario of its origin; 2012 VP$_{113}$ requires $Q_{2}>6\times 10^{-27}$ s$^{-2}$ , which is more than $1\sigma$ off of the nominal value of \cite{H+14}, or $Q_{2}<4\times 10^{-27}$ s$^{-2}$ , which is more than $2.3\sigma$ off of the nominal value of \cite{H+14}, to securely state that its perihelion was lifted purely by EFE.

\item Similar to the P9 scenario, EFE does not induce strong confinement in $\omega-a$ plane until $a\gtrsim 600$ au, the clustering in $\omega$ is traced only by the potentially unstable orbits, and the inferred distribution of $\omega$ at this $a$-range is bimodal (peaking around $\omega=0$ and 180 deg); moreover, the confinement fades away when $Q_{2}$ is not close to $Q_{2}=1\times 10^{-27}$ s$^{-2}$.

\item Concerning perihelion longitude $L$, visually appealing clustering traced by the potentially unstable orbits develops for $a\gtrsim 300$ au and wide range of values of $Q_{2}$; if $Q_{2}<0$, the clustering centres close to 90 and 270 deg, while when $Q_{2}>0$, the clustering centres close to 0 (360) and 180 deg.

\item In the stable population, the clustering in $L$ looms for $\vert Q_{2}\vert\gtrsim 4\times 10^{-27}$ s$^{-2}$, with distribution of $L$ peaking close to 90 and 270 deg.

\item The clustering in longitude of ascending node $\Omega$ is the most visually appealing and robust,  spanning wide range of permitted $Q_{2}$ values, and traced by both stable and potentially unstable orbits with $a>150$ au; depending on the sign of $Q_{2}$, either $\Omega\in(0,180)$ deg $(Q_{2}<0)$ or $\Omega\in(180,360)$ deg $(Q_{2}>0)$ is preferred by the simulated ETNOs, especially when we look at $a>250$ au orbits.

\item If $\vert Q_{2}\vert\gtrsim4\times 10^{-27}$ s$^{-2}$, $a>250$ au ETNOs with very low inclinations are suppressed, while the population with inclinations between 10 and 30 deg is relatively enhanced.

\item High inclination and retrograde orbits of large semimajor axis Centaurs can be produced by EFE.
\end{itemize}

These results are in some tension with the present-day observations. First of all, the simulations did not reproduce the tightness of the clustering in $\omega$ down to $a=150$ au. Second, the clustering region in $\Omega$ is shifted by 180 deg with respect to the observed region, when one assumes, as usual, positive $Q_{2}$. There are not any interpolating function families known, giving arise to $Q_{2}<0$ in AQUAL or QUMOND at the present time, though maybe these are possible to construct.
Assuming $Q_{2}$ to be negative solves the problem. For instance $Q_{2}=-1\times 10^{-27}$ s$^{-2}$ (1.3$\sigma$ off of the nominal $Q_{2}$ of \cite{H+14}) yields clustering in $\Omega$ in line with the observations. The observed clustering in $\Omega$ (if it is of dynamical origin) constrains $Q_{2}$ to $Q_{2}<0$. Perihelia of Sednoids and ETNOs distributions of $L$ and $i$ prefer $\vert Q_{2}\vert >4\times 10^{-27}$ s$^{-2}$. The $Q_{2}$s lower than $Q_{2}=-4\times 10^{-27}$ s$^{-2}$ (2.3$\sigma$ off of the nominal $Q_{2}$ of \cite{H+14}) are very unlikely, with a probability of about $1\%,$ in  light of the Cassini data.

Our results predict the effect MD could have on ETNOs and should be compared with future unbiased and robust data, when they are acquired. This means further constraints on the interpolating function as it is directly linked to the value of $Q_{2}$.
Also, provided that P9 were found, $Q_{2}$ and the interpolating function would be further constrained in this case.
Unfortunately such a test is not of the Popper's standard of the hypotheses testing. We will not be able to rule out MD completely this way. It is always possible to construct interpolating function giving unmeasurable EFE. Moreover, there are flavours of MD completely without any EFE.

Does P9 do the job better? 
With P9, one has more freedom to fit the data. Planet 9 in an eccentric orbit can be made massive, hence its effect on ETNOs is strong, while it can be hidden close to its aphelion at the present time in order not to violate the Cassini data or planetary ephemerides\footnote{\cite{Fie+16} found that the true anomaly, $\phi$, most probably lies in the interval $(108,129)$ deg, using the same orbit of P9 as we considered in this paper.}. There is much less freedom in the MD scenario. The external field direction rotates uniformly and very slowly (with period of one Galactic year) around the Sun, while $Q_{2}$, the only free parameter, stays approximately constant and bounded from above with aid of the Cassini data.
The clustering in $\omega$ seems problematic in both frameworks; with the tuned orbit of Planet 9, $\omega$ is still not confined until $a\gtrsim 500$ au; see our Fig. \ref{img:p91_om} or Fig. 8 in \cite{BB16}. Moreover, the results of \cite{S+16} and \cite{Law+17} pose a problem for the P9 scenario, as already discussed above. It is unclear whether similar issues would also arise  in the scenario where EFE, not P9, shapes the orbits of ETNOs.

In this paper we applied the Occam's razor approach; we tried to explain the clustered orbits of ETNOs with the effect of MD only. Maybe a more complicated dynamical model, where a distant planetary perturber exists while EFE of MD is also relevant, could perform better in explaining ETNOs orbits, the overall structure of the Kuiper belt, and the origin of the distant planet at the same time.

\begin{acknowledgements}
I am thankful to Jozef Vil\'{a}gi and Leonard Korno\v{s} for providing a computational facility. I also thank Vlad\'{i}mir Balek, Jozef Kla\v{c}ka, and Leonard Korno\v{s} for discussions and Aur\'{e}lien Hees for useful correspondence. This work was supported by Comenius University Grant no. UK/419/2016.
\end{acknowledgements}

\bibliographystyle{aa}
\bibliography{clustering}

\begin{thebibliography}{67}
\expandafter\ifx\csname natexlab\endcsname\relax\def\natexlab#1{#1}\fi

\bibitem[{{Batygin} \& {Brown}(2016)}]{BB16}
{Batygin}, K. \& {Brown}, M.~E. 2016, \aj, 151, 22

\bibitem[{{Bekenstein} \& {Milgrom}(1984)}]{BM84}
{Bekenstein}, J. \& {Milgrom}, M. 1984, \apj, 286, 7

\bibitem[{{Bekenstein}(2004)}]{Bek04}
{Bekenstein}, J.~D. 2004, \prd, 70, 083509

\bibitem[{{Blanchet} \& {Novak}(2011)}]{BN11}
{Blanchet}, L. \& {Novak}, J. 2011, \mnras, 412, 2530

\bibitem[{{Bosma}(1981)}]{Bos81}
{Bosma}, A. 1981, \aj, 86, 1825

\bibitem[{{Boylan-Kolchin} {et~al.}(2012){Boylan-Kolchin}, {Bullock}, \&
  {Kaplinghat}}]{MBK+12}
{Boylan-Kolchin}, M., {Bullock}, J.~S., \& {Kaplinghat}, M. 2012, \mnras, 422,
  1203

\bibitem[{{Brasser} {et~al.}(2012){Brasser}, {Duncan}, {Levison}, {Schwamb}, \&
  {Brown}}]{Bra+12}
{Brasser}, R., {Duncan}, M.~J., {Levison}, H.~F., {Schwamb}, M.~E., \& {Brown},
  M.~E. 2012, \icarus, 217, 1

\bibitem[{{Brown}(2001)}]{Bro01}
{Brown}, M.~E. 2001, \aj, 121, 2804

\bibitem[{{Brown} \& {Batygin}(2016)}]{BB16b}
{Brown}, M.~E. \& {Batygin}, K. 2016, \apjl, 824, L23

\bibitem[{{Brown} {et~al.}(2004){Brown}, {Trujillo}, \& {Rabinowitz}}]{Bro+04}
{Brown}, M.~E., {Trujillo}, C., \& {Rabinowitz}, D. 2004, \apj, 617, 645

\bibitem[{{Brown} \& {Firth}(2016)}]{BF16}
{Brown}, R.~B. \& {Firth}, J.~A. 2016, \mnras, 456, 1587

\bibitem[{{Caldwell} {et~al.}(2016){Caldwell}, {Walker}, {Mateo}, {Olszewski},
  {Koposov}, {Belokurov}, {Torrealba}, {Geringer-Sameth}, \& {Johnson}}]{C+16}
{Caldwell}, N., {Walker}, M.~G., {Mateo}, M., {et~al.} 2016, ArXiv e-prints
  [\eprint[arXiv]{1612.06398}]

\bibitem[{{Chambers}(1999)}]{Cha99}
{Chambers}, J.~E. 1999, \mnras, 304, 793

\bibitem[{{Cowan} {et~al.}(2016){Cowan}, {Holder}, \& {Kaib}}]{Cow+16}
{Cowan}, N.~B., {Holder}, G., \& {Kaib}, N.~A. 2016, \apjl, 822, L2

\bibitem[{{de la Fuente Marcos} \& {de la Fuente Marcos}(2014)}]{dlFdlF14}
{de la Fuente Marcos}, C. \& {de la Fuente Marcos}, R. 2014, \mnras, 443, L59

\bibitem[{{de la Fuente Marcos} \& {de la Fuente Marcos}(2016)}]{dlFdlF16}
{de la Fuente Marcos}, C. \& {de la Fuente Marcos}, R. 2016, \mnras, 459, L66

\bibitem[{{de la Fuente Marcos} {et~al.}(2016){de la Fuente Marcos}, {de la
  Fuente Marcos}, \& {Aarseth}}]{dlF+16}
{de la Fuente Marcos}, C., {de la Fuente Marcos}, R., \& {Aarseth}, S.~J. 2016,
  \mnras, 460, L123

\bibitem[{{Desmond}(2017)}]{Des16}
{Desmond}, H. 2017, \mnras, 464, 4160

\bibitem[{{Dones} {et~al.}(2015){Dones}, {Brasser}, {Kaib}, \&
  {Rickman}}]{D+15}
{Dones}, L., {Brasser}, R., {Kaib}, N., \& {Rickman}, H. 2015, \ssr, 197, 191

\bibitem[{{Einstein}(1916)}]{E1916}
{Einstein}, A. 1916, Annalen der Physik, 354, 769

\bibitem[{{Famaey} \& {McGaugh}(2012)}]{FM12}
{Famaey}, B. \& {McGaugh}, S.~S. 2012, Living Reviews in Relativity, 15, 10

\bibitem[{{Fienga} {et~al.}(2016){Fienga}, {Laskar}, {Manche}, \&
  {Gastineau}}]{Fie+16}
{Fienga}, A., {Laskar}, J., {Manche}, H., \& {Gastineau}, M. 2016, \aap, 587,
  L8

\bibitem[{{Fortney} {et~al.}(2016){Fortney}, {Marley}, {Laughlin},
  {Nettelmann}, {Morley}, {Lupu}, {Visscher}, {Jeremic}, {Khadder}, \&
  {Hargrave}}]{For+16}
{Fortney}, J.~J., {Marley}, M.~S., {Laughlin}, G., {et~al.} 2016, \apjl, 824,
  L25

\bibitem[{{Gomes} {et~al.}(2006){Gomes}, {Matese}, \& {Lissauer}}]{Gom+06}
{Gomes}, R.~S., {Matese}, J.~J., \& {Lissauer}, J.~J. 2006, \icarus, 184, 589

\bibitem[{{Gomes} {et~al.}(2015){Gomes}, {Soares}, \& {Brasser}}]{Gom+15}
{Gomes}, R.~S., {Soares}, J.~S., \& {Brasser}, R. 2015, \icarus, 258, 37

\bibitem[{{Gulbis} {et~al.}(2010){Gulbis}, {Elliot}, {Adams}, {Benecchi},
  {Buie}, {Trilling}, \& {Wasserman}}]{Gul+10}
{Gulbis}, A.~A.~S., {Elliot}, J.~L., {Adams}, E.~R., {et~al.} 2010, \aj, 140,
  350

\bibitem[{{Haghi} {et~al.}(2016){Haghi}, {Bazkiaei}, {Zonoozi}, \&
  {Kroupa}}]{Hag+16}
{Haghi}, H., {Bazkiaei}, A.~E., {Zonoozi}, A.~H., \& {Kroupa}, P. 2016, \mnras,
  458, 4172

\bibitem[{{Hees} {et~al.}(2016){Hees}, {Famaey}, {Angus}, \& {Gentile}}]{H+16}
{Hees}, A., {Famaey}, B., {Angus}, G.~W., \& {Gentile}, G. 2016, \mnras, 455,
  449

\bibitem[{{Hees} {et~al.}(2014){Hees}, {Folkner}, {Jacobson}, \& {Park}}]{H+14}
{Hees}, A., {Folkner}, W.~M., {Jacobson}, R.~A., \& {Park}, R.~S. 2014, \prd,
  89, 102002

\bibitem[{{Holman} \& {Payne}(2016{\natexlab{a}})}]{HP16b}
{Holman}, M.~J. \& {Payne}, M.~J. 2016{\natexlab{a}}, \aj, 152, 80

\bibitem[{{Holman} \& {Payne}(2016{\natexlab{b}})}]{HP16}
{Holman}, M.~J. \& {Payne}, M.~J. 2016{\natexlab{b}}, \aj, 152, 94

\bibitem[{{Iorio}(2010)}]{Ior10}
{Iorio}, L. 2010, The Open Astronomy Journal, 3, 1

\bibitem[{{Iorio}(2013)}]{Ior13}
{Iorio}, L. 2013, Classical and Quantum Gravity, 30, 165018

\bibitem[{{J{\'{\i}}lkov{\'a}} {et~al.}(2015){J{\'{\i}}lkov{\'a}}, {Portegies
  Zwart}, {Pijloo}, \& {Hammer}}]{Jil+15}
{J{\'{\i}}lkov{\'a}}, L., {Portegies Zwart}, S., {Pijloo}, T., \& {Hammer}, M.
  2015, \mnras, 453, 3157

\bibitem[{{Kaib} \& {Quinn}(2008)}]{KQ08}
{Kaib}, N.~A. \& {Quinn}, T. 2008, \icarus, 197, 221

\bibitem[{{Kaib} {et~al.}(2011){Kaib}, {Ro{\v s}kar}, \& {Quinn}}]{Kai+11}
{Kaib}, N.~A., {Ro{\v s}kar}, R., \& {Quinn}, T. 2011, \icarus, 215, 491

\bibitem[{{Lawler} {et~al.}(2017){Lawler}, {Shankman}, {Kaib}, {Bannister},
  {Gladman}, \& {Kavelaars}}]{Law+17}
{Lawler}, S.~M., {Shankman}, C., {Kaib}, N., {et~al.} 2017, \aj, 153, 33

\bibitem[{{Le Verrier}(1846)}]{LeV46}
{Le Verrier}, U.~J. 1846, Astronomische Nachrichten, 25, 65

\bibitem[{{Le Verrier}(1859)}]{LeV59}
{Le Verrier}, U.~J. 1859, Annales de l'Observatoire de Paris, 5

\bibitem[{{Lelli} {et~al.}(2016){Lelli}, {McGaugh}, {Schombert}, \&
  {Pawlowski}}]{L+16}
{Lelli}, F., {McGaugh}, S.~S., {Schombert}, J.~M., \& {Pawlowski}, M.~S. 2016,
  \apjl, 827, L19

\bibitem[{{Lelli} {et~al.}(2017){Lelli}, {McGaugh}, {Schombert}, \&
  {Pawlowski}}]{L+17}
{Lelli}, F., {McGaugh}, S.~S., {Schombert}, J.~M., \& {Pawlowski}, M.~S. 2017,
  \apj, 836, 152

\bibitem[{{McGaugh} \& {Milgrom}(2013{\natexlab{a}})}]{McGM13a}
{McGaugh}, S. \& {Milgrom}, M. 2013{\natexlab{a}}, \apj, 766, 22

\bibitem[{{McGaugh} \& {Milgrom}(2013{\natexlab{b}})}]{McGM13b}
{McGaugh}, S. \& {Milgrom}, M. 2013{\natexlab{b}}, \apj, 775, 139

\bibitem[{{McGaugh}(2004)}]{McG04}
{McGaugh}, S.~S. 2004, \apj, 609, 652

\bibitem[{{McGaugh}(2016)}]{McG16}
{McGaugh}, S.~S. 2016, \apjl, 832, L8

\bibitem[{{McGaugh} {et~al.}(2016){McGaugh}, {Lelli}, \& {Schombert}}]{McG+16}
{McGaugh}, S.~S., {Lelli}, F., \& {Schombert}, J.~M. 2016, Physical Review
  Letters, 117, 201101

\bibitem[{{McGaugh} {et~al.}(2000){McGaugh}, {Schombert}, {Bothun}, \& {de
  Blok}}]{M+00}
{McGaugh}, S.~S., {Schombert}, J.~M., {Bothun}, G.~D., \& {de Blok}, W.~J.~G.
  2000, \apjl, 533, L99

\bibitem[{{McMillan} \& {Binney}(2010)}]{McM+10}
{McMillan}, P.~J. \& {Binney}, J.~J. 2010, \mnras, 402, 934

\bibitem[{{Milgrom}(1983{\natexlab{a}})}]{Mil83a}
{Milgrom}, M. 1983{\natexlab{a}}, \apj, 270, 371

\bibitem[{{Milgrom}(1983{\natexlab{b}})}]{Mil83b}
{Milgrom}, M. 1983{\natexlab{b}}, \apj, 270, 365

\bibitem[{{Milgrom}(1994)}]{Mil94}
{Milgrom}, M. 1994, Annals of Physics, 229, 384

\bibitem[{{Milgrom}(2009{\natexlab{a}})}]{Mil09}
{Milgrom}, M. 2009{\natexlab{a}}, \mnras, 399, 474

\bibitem[{{Milgrom}(2009{\natexlab{b}})}]{Mil09b}
{Milgrom}, M. 2009{\natexlab{b}}, \apj, 698, 1630

\bibitem[{{Milgrom}(2010)}]{Mil10}
{Milgrom}, M. 2010, \mnras, 403, 886

\bibitem[{{Milgrom}(2012)}]{Mil12}
{Milgrom}, M. 2012, \mnras, 426, 673

\bibitem[{{Milgrom}(2014)}]{Mil14}
{Milgrom}, M. 2014, \mnras, 437, 2531

\bibitem[{{Milgrom}(2016{\natexlab{a}})}]{Mil16}
{Milgrom}, M. 2016{\natexlab{a}}, ArXiv e-prints [\eprint[arXiv]{1605.07458}]

\bibitem[{{Milgrom}(2016{\natexlab{b}})}]{Mil16b}
{Milgrom}, M. 2016{\natexlab{b}}, Physical Review Letters, 117, 141101

\bibitem[{{Morbidelli} \& {Levison}(2004)}]{ML04}
{Morbidelli}, A. \& {Levison}, H.~F. 2004, \aj, 128, 2564

\bibitem[{{Pau{\v c}o} \& {Kla{\v c}ka}(2016)}]{PK16}
{Pau{\v c}o}, R. \& {Kla{\v c}ka}, J. 2016, \aap, 589, A63

\bibitem[{{Rubin} {et~al.}(1982){Rubin}, {Ford}, {Thonnard}, \&
  {Burstein}}]{R+82}
{Rubin}, V.~C., {Ford}, Jr., W.~K., {Thonnard}, N., \& {Burstein}, D. 1982,
  \apj, 261, 439

\bibitem[{{Sancisi}(2004)}]{Ren04}
{Sancisi}, R. 2004, in IAU Symposium, Vol. 220, Dark Matter in Galaxies, ed.
  S.~{Ryder}, D.~{Pisano}, M.~{Walker}, \& K.~{Freeman}, 233

\bibitem[{{Sanders}(1990)}]{San90}
{Sanders}, R.~H. 1990, \aapr, 2, 1

\bibitem[{{Sch{\"o}nrich}(2012)}]{Sch12}
{Sch{\"o}nrich}, R. 2012, \mnras, 427, 274

\bibitem[{{Shankman} {et~al.}(2017){Shankman}, {Kavelaars}, {Lawler},
  {Gladman}, \& {Bannister}}]{S+16}
{Shankman}, C., {Kavelaars}, J.~J., {Lawler}, S.~M., {Gladman}, B.~J., \&
  {Bannister}, M.~T. 2017, \aj, 153, 63

\bibitem[{{Sheppard} \& {Trujillo}(2016)}]{ST16}
{Sheppard}, S.~S. \& {Trujillo}, C. 2016, \aj, 152, 221

\bibitem[{{Trujillo} \& {Sheppard}(2014)}]{TS14}
{Trujillo}, C.~A. \& {Sheppard}, S.~S. 2014, \nat, 507, 471

\end{thebibliography}

\end{document}